\documentclass[12pt,preprint]{aastex}
\shorttitle{ChaMP Galaxy Cluster Survey}
\shortauthors{Barkhouse et al.}
\begin{document}
\title{ChaMP Serendipitous Galaxy Cluster Survey}
\author{W. A. Barkhouse\altaffilmark{1,2}, P. J. Green\altaffilmark{1,2}, 
A. Vikhlinin\altaffilmark{2}, D.-W. Kim\altaffilmark{2}, D. Perley
\altaffilmark{3}, R. Cameron\altaffilmark{4}, J. Silverman\altaffilmark{1,5}, 
A. Mossman\altaffilmark{2}, R. Burenin\altaffilmark{6}, B. T. Jannuzi
\altaffilmark{7}, M. Kim\altaffilmark{2}, M. G. Smith\altaffilmark{8}, 
R. C. Smith\altaffilmark{8}, H. Tananbaum\altaffilmark{2}, and B. J. 
Wilkes\altaffilmark{2}\altaffiltext{1}{Visiting Astronomer, Cerro Tololo 
Inter-American Observatory and Kitt Peak National Observatory, National 
Optical Astronomy Observatory, operated by the Association for Universities 
for Research in Astronomy, Inc. (AURA), under contract to the National 
Science Foundation.}
\altaffiltext{2}{Harvard-Smithsonian, Center for Astrophysics, 60 Garden 
Street, Cambridge, MA 02138.}
\altaffiltext{3}{Department of Astronomy, 601 Campbell Hall, University of
California at Berkeley, Berkeley, CA 94720.}
\altaffiltext{4}{Stanford Linear Accelerator Center, 2575 Sand Hill Road, 
Menlo Park, CA 94025.}
\altaffiltext{5}{Max-Plank-Institut f\"ur extraterrestrische Physik, 
Giessenbachstrasse, D-85741 Garching, Germany.}
\altaffiltext{6}{Space Research Institute, Moscow, Russia.}
\altaffiltext{7}{National Optical Astronomy Observatory, P.O. Box 26732, 
Tucson, AZ 85726.}
\altaffiltext{8}{Cerro Tololo Inter-American Observatory, National Optical 
Astronomy Observatory, Casilla 603, La Serena, Chile.}
}

\begin{abstract}

We present a survey of serendipitous extended X-ray sources and optical 
cluster candidates from the {\it Chandra} Multi-wavelength Project (ChaMP). 
Our main goal is to make an unbiased comparison of X-ray and optical 
cluster detection methods. In 130 archival {\it Chandra} pointings covering 
13 square degrees, we use a wavelet decomposition technique to detect 55 
extended sources, of which 6 are nearby single galaxies. Our X-ray cluster 
catalog reaches a typical flux limit of about 
$\sim10^{-14}$~erg~s$^{-1}$~cm$^{-2}$, with a median cluster core 
radius of 21\arcsec. For 56 of the 130 X-ray fields, we use the ChaMP's 
deep NOAO/4m MOSAIC $g^{\prime}$, $r^{\prime}$, and $i^{\prime}$ imaging 
to independently detect cluster candidates using a Voronoi tessellation 
and percolation (VTP) method. Red-sequence filtering decreases the 
galaxy fore/background contamination and provides photometric redshifts 
to $z\sim0.7$. From the overlapping 6.1 square degree X-ray/optical imaging, 
we find 115 optical clusters (of which 11\% are in the X-ray catalog) and 
28 X-ray clusters (of which 46\% are in the optical VTP catalog ). 
The median redshift of the 13 X-ray/optical clusters is 
0.41, and their median X-ray luminosity (0.5-2 keV) is 
$L_X=(2.65\pm0.19)\times10^{43}~\mbox{ergs}~\mbox{s}^{-1}$. The clusters 
in our sample that are only detected in our optical data are poorer on 
average ($\sim 4\sigma$) than the X-ray/optically matched clusters, which 
may partially explain the difference in the detection fractions.

\end{abstract}


\section{Introduction}

A primary goal of modern astronomy is to study the formation and evolution 
of galaxies. Clusters of galaxies provide us with laboratories where 
galaxy evolution can be studied over a large range in cosmic look-back 
time. The high-density cluster environment probes the impact of high 
galaxy density on the fate of the cluster galaxy population. Interactions, 
mergers, and dynamical effects (e.g., tidal forces and ram pressure 
stripping) may play significant roles in shaping galaxy evolution 
in these type of locales \citep[e.g.,][]{Dub98,Moore99}.

Galaxy clusters are also the most massive, mainly virialized, 
concentrations of matter in the Universe and act as tracers of the 
underlying dark matter. Clusters thus also play a key role constraining 
fundamental cosmological parameters such as $\Omega_{m}$ (the matter-density 
parameter) and $\sigma_{8}$ (the root-mean-square density fluctuation on 
a scale of $8h^{-1}~\mbox{Mpc}$). The number density of clusters as a 
function of mass and redshift strongly depends on $\Omega_{m}$ and 
$\sigma_{8}$ (see Rosati, Borgani, \& Norman 2002, and references therein). 
This remarkable feature of hierarchical cluster formation, via the 
Press-Schechter formalism \citep{Press74}, affords the opportunity to 
provide an independent confirmation of various cosmological quantities 
obtained recently by other techniques \citep[e.g., WMAP;][]{Bennett03}.

To facilitate the investigation of galaxy cluster evolution and 
provide constraints on cosmological parameters, a large sample of 
galaxy clusters spanning a redshift range from $0 < z < 1.5$ is 
required. The search for galaxy clusters has been conducted mainly 
using optical and X-ray techniques. Although these methods overlap 
in their ability to distinguish galaxy clusters from the general 
background, they sample different regions of parameter space that 
encompass nearly the complete range in physical attributes of clusters. 
For example, X-ray techniques detect clusters via extended emission from 
the hot gas that makes up the intracluster medium 
\citep[ICM; e.g.,][]{Vikhlinin98b}. X-ray detection suffers little from 
source confusion \citep[e.g.,][]{Basilakos04,Popesso04}, but tends to 
select more massive, virialized clusters. By contrast, optical methods 
rely on the detection of an overdensity of galaxies or a population of 
early-type galaxies with a narrow range of colors 
\citep[red-sequence;][]{Gladd00}. Optical methods are generally more 
sensitive to non-virialized (or young) systems than X-ray searches, 
but are also susceptible to projection effects, and bias toward more 
evolved galaxy populations. Multi-wavelength cluster detection schemes 
help to insure a higher degree of completeness and reliability in any 
cluster compilation \citep{Postman02a}. 

Several recent studies have compared cluster samples compiled from 
independent techniques using optical and X-ray data. \citet{Donahue02}, 
for example, applied a matched-filter method to optical data and found 
that 60\% (26 out of 43) of {\it ROSAT} X-ray clusters had optical 
matches. They also determined that optical clusters/groups outnumbered 
X-ray extended sources by a factor of three. \citet{Gilbank04}, applying 
a matched-filter algorithm to optical data, found 75\% matches (9/12) to 
a sample of {\it ROSAT} X-ray clusters. Using a cluster detection method 
based on the red-sequence of early-type cluster galaxies, \citet{Gilbank04} 
was able to achieve a matched fraction of 100\% (10/10) using the same 
X-ray dataset. In addition, \citet{Basilakos04} and \citet{Kolokotronis06} 
--- using a smoothing percolation technique on optical data --- found 
matched fractions of 75\% (3/4) and 68\% (13/19), respectively, 
for extended X-ray sources compiled from archival {\it XMM-Newton} 
observations. In all such studies, the matched fractions depend on the 
relative optical/X-ray flux limits and the sensitivity of the detection 
algorithms.

The aims of this paper are to present a new serendipitous X-ray cluster 
sample based on data from the {\it Chandra} Multi-wavelength Project 
\citep[ChaMP;][see http://hea-www.harvard.edu/CHAMP/]{Kim04a,Green04}, 
and to make an unbiased comparison of X-ray and optical cluster detection 
methods. We thus explore a variety of questions: Are there massive, X-ray 
luminous, clusters that are optically poor? Do all massive clusters emit 
X-rays? What types of optical clusters retain hot gas and why? In addition, 
we present the X-ray and optical properties of our sample of 
serendipitously-detected clusters/groups, including a comparison of X-ray 
luminosity with optical cluster richness. Finally, we provide the community 
with a compilation of newly-discovered clusters/groups 
that can be used in conjunction with other samples to constrain cosmological 
parameters. This sample should also help to address how the more numerous 
(but less well-studied) low-luminosity clusters and groups evolve.

This paper is organized as follows. In \S 2 we describe the sample selection 
and X-ray and optical observations. In \S 3 we discuss our X-ray and optical 
cluster detection methods. The properties of our X-ray and optical cluster 
candidates are presented in \S 4, along with a comparison of the two 
compilations. Finally, in \S 5 we compare our results with previous studies 
and discuss possible bias inherit in our X-ray and optical cluster detection 
schemes. Unless otherwise indicated, we use $\Omega_{\mbox{m}}= 0.3$, 
$\Omega_{\lambda}= 0.7$, and 
$\mbox{H}_{\mbox{o}}= 70~\mbox{km}~\mbox{s}^{-1}~\mbox{Mpc}^{-1}$ throughout.

\section{Sample Selection and Observations}

This study makes use of the data provided by the ChaMP. The ChaMP is a 
$\sim 13~\mbox{deg}^{2}$ (based on Cycles 1 \& 2 {\it Chandra} archival data) 
survey of serendipitous {\it Chandra} X-ray sources at flux 
levels ($f_X\sim 10^{-15}-10^{-14}~\mbox{erg}~\mbox{sec}
^{-1}~\mbox{cm}^{-2}$), intermediate between the {\it Chandra} Deep surveys 
and previous X-ray missions. Optical follow-up of ChaMP fields was conducted 
using the MOSAIC camera on the KPNO and CTIO 4m telescopes. The mosaic 
imaging of ChaMP was designed to search for optical counterparts to active 
galactic nuclei (AGNs) in part to drive our spectroscopy identification 
program. Presently, 56 mosaic fields in $g^{\prime}$, $r^{\prime}$, 
and $i^{\prime}$ to $r^{\prime}\la 25$ ($5\sigma$ detection) have been 
acquired (Barkhouse et al. in preparation). For a description of ChaMP 
methods, analysis and early science results, see Kim et al. 
(2004a, 2004b, 2005), Green et al. (2004), Silverman (2004), and 
Silverman et al. (2005a, 2005b). 

The X-ray data for this study are drawn from 130 fields selected from 
{\it Chandra} AO1 and AO2 observing periods. The fields were selected 
based on the following criteria; i) include only ACIS imaging fields 
(excluding the ACIS-S4 chip), ii) include only fields more than $20\degr$ 
from the Galactic plane to minimize extinction, iii) exclude fields 
dominated by large extended sources, iv) no planetary observations, v) no 
survey observations by PI, and vi) no fields close to the LMC, SMC, and 
M31 (see Kim et al. 2004a for a detailed discussion of selection criteria 
and X-ray data reductions.) Galactic $\mbox{N}_{\mbox{H}}$ values are taken 
from \cite{Stark92} and are tabulated in Table 1 for fields containing at 
least one extended X-ray source.

The optical data for this study consists of ChaMP mosaic images acquired 
from NOAO 4m telescopes in the $g^{\prime}$, $r^{\prime}$, and $i^{\prime}$ 
bandpasses (see Table 2). The optical and X-ray imaging overlap by 6.1 square 
degrees. These data are used for source identification and to compare 
optical cluster detection methods against X-ray techniques for the area in 
common (see \S 3 below). Details of image 
reduction and analysis for the initial sample of six ChaMP mosaic fields 
are presented in \cite{Green04}, and an overview of our complete sample 
of 56 fields in Barkhouse et al. (in preparation). In summary, our optical 
exposure times were scaled to the X-ray exposures to probe a constant 
X-ray/optical flux ratio. The optical follow-up was optimized to probe 
AGN counterparts and not faint galaxies at a similar redshift for a given 
X-ray luminosity. The image reduction was performed using the 
{\tt mscred} package within the IRAF\footnote{IRAF is distributed by the 
National Optical Astronomy Observatory, which is operated by the Association 
of Universities for Research in Astronomy, Inc., under the cooperative 
agreement with the National Science Foundation.} environment. Object 
detection and photometry was conducted using SExtractor \citep{Bertin96}. 
Photometric calibrations were done using standard stars from 
\citet{Landolt92}, which were converted to the SDSS photometric system 
using the transformation equations from \cite{Fukugita96}. Table 2 
summarizes the optical properties of the 36 mosaic fields that 
overlap X-ray fields containing X-ray-detected extended sources or 
optical cluster candidates.

\section{Galaxy Cluster Detection Methods}

We have employed the X-ray and optical datasets from ChaMP to search for and 
contrast galaxy cluster samples compiled independently from X-ray and optical 
cluster search techniques for the overlapping 6.1 square degree sky coverage. 
In the following sections we describe each 
detection method with an emphasis on the description of the optical 
technique (see Vikhlinin et al. 1998b for a detailed description of the
extended X-ray source detection algorithm). 

\subsection{Extended X-ray Source Detection}

The extended X-ray source detection is based on a wavelet decomposition 
technique --- plus a maximum-likelihood method to determine the 
significance of each detected extended source --- that is similar to the 
method described in \citet{Vikhlinin98b}. In brief, each extended source 
was detected in the 0.7-2 keV energy band to maximize the contrast of the 
cluster ICM against the X-ray background. A Gaussian kernel was fit to each 
wavelet source and its best-fit radius was compared with the PSF size 
appropriate for the measured off-axis angle. Those objects determined to be 
``point-like'' were then subtracted and the detection process applied to 
the resultant image. The sample of X-ray sources deemed ``extended'' were 
then fit on the original image to a standard $\beta$-model, 
$I(r,r_{c})=I_{0}[1+(r/r_{c})^{2}]^{-3\beta + 0.5}$ \citep{Cavaliere76}, 
with point sources masked out. Since a free fit 
was not possible due to the small number of photons expected for most 
sources, we fixed the value for $\beta$ at $0.67$ 
\citep[e.g.,][]{Vikhlinin98b,Donahue02,Moretti04}.  

\subsubsection{Final X-ray-Selected Cluster Catalog}

\begin{figure}
\figurenum{1}
\epsscale{0.9}
\plotone{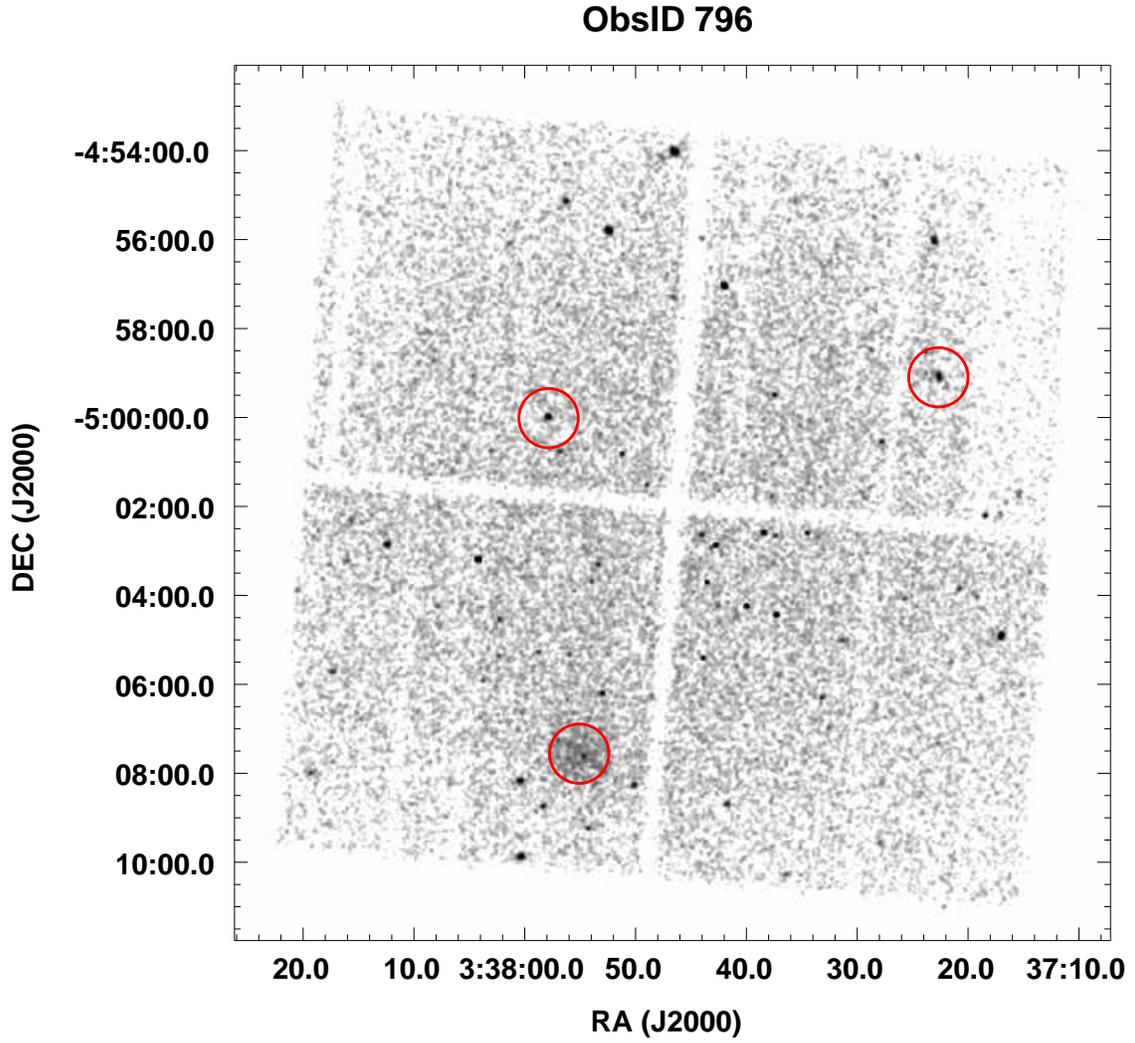}
\caption{Part of a smoothed {\it Chandra} image, ObsID 796, showing the 
location of three serendipitously-detected extended X-ray sources on three 
ACIS-I chips (circles; $40\arcsec$ in radius). The PI target, the blue 
compact dwarf galaxy SBS 0335-052, is located near the center of the image.}
\label{ObsID796}
\end{figure}

The initial sample of extended X-ray sources is comprised of PI target 
clusters, serendipitous clusters, nearby bright galaxies, and spurious 
detections caused by chip gaps, edge effects, etc. Visual inspection 
and cross-correlation to {\it Chandra} PI targets was used to assemble 
a final list consisting of 55 high-confidence serendipitously-detected 
extended sources (see Table 1 and Figure~\ref{ObsID796}). From the 
sample of 55 extended X-ray sources, 6 were found to be associated with 
low-redshift galaxies (three ellipticals, two spirals, and an S0/Sa galaxy).

The X-ray flux for each source was computed from the total number of 
counts by extrapolating the $\beta$-model fit to infinity. Also assumed 
was a Raymond-Smith thermal spectrum with a temperature of $T_X=2$ keV, a 
solar abundance of $Z_{\sun}=0.3$, and galactic extinction appropriate for 
each field. We use $T_X=2$ keV since it is appropriate based on the median 
$L_X$ of our cluster sample ($L_X\sim 10^{43}~\mbox{erg}~\mbox{s}^{-1}$) 
and the $T_X$-$L_X$ relation \citep[e.g.,][]{White97}. Using $T_X=5$ keV, 
for example, will change $f_X$ by $\sim 8\%$. 

\begin{figure}
\figurenum{2}
\includegraphics[angle=-90,scale=0.8]{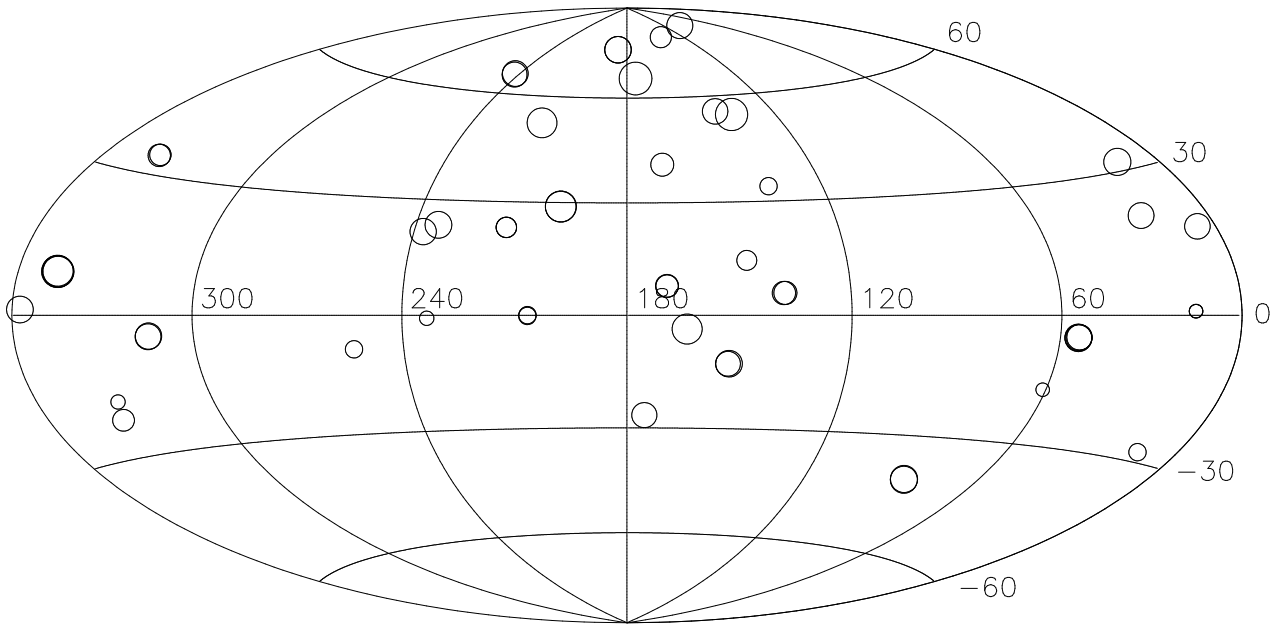}
\caption{Aitoff equatorial projection in galactic coordinates of 55 extended 
X-ray sources. Symbol size is proportional to X-ray exposure time 
(see Table 1).}
\label{Aitoff}
\end{figure}

X-ray flux values were converted to the 0.5-2 keV energy 
band and uncertainties derived from Poisson statistics. X-ray luminosities 
were calculated from measured fluxes using redshift estimates derived from 
(in order of preference); 1) the ChaMP spectroscopic program \citep{Green04}, 
2) published spectroscopic redshifts, or 3) estimated from our 
red-sequence filtered VTP optical cluster detection method (see \S 3.2). In 
Table 3 the X-ray properties of our extended source catalog are tabulated. 
Figure~\ref{Aitoff} shows the all-sky distribution of our final sample of 
55 extended X-ray sources.

\subsection{Optical Cluster Detection}

The detection of galaxy clusters from optical data has had a long history 
dating back to the pioneering work of \citet{Abell58} and \citet{Zwicky61}. 
Various techniques have been employed to take advantage of the expected 
shape, luminosity function, density enhancement, and color distribution of the 
cluster galaxy population. Some of the automated methods include; the 
matched-filter algorithm \citep {Postman96}, the Voronoi tessellation 
percolation method \citep{Ebeling93,Ramella01}, the cluster red-sequence 
technique \citep{Gladd00}, the detection of surface brightness fluctuations 
\citep{Gonz01}, and the maxBCG procedure \citep{Bahcall03}. The properties 
of the resulting cluster samples naturally differ depending on the data 
quality, the detection technique and selection criteria 
\citep[e.g.,][]{Donahue02,Kim02}.  

\subsubsection{Red-Sequence Voronoi Tessellation Percolation Algorithm}

The construction of a cluster catalog from ChaMP optical images is based 
on a modified version of the Voronoi tessellation and percolation 
technique (VTP) described in \citet{Ebeling93} and 
\citet{Ramella01}\footnote{The VTP code was downloaded from 
http://www.ts.astro.it/astro/VoroHome/.}. The advantage of implementing 
this type of non-parametric algorithm is that no assumption is made 
regarding cluster shapes --- as is the case for the matched-filter code --- 
and thus VTP is sensitive to irregular clusters as well as symmetric ones.

\begin{figure}
\figurenum{3}
\plotone{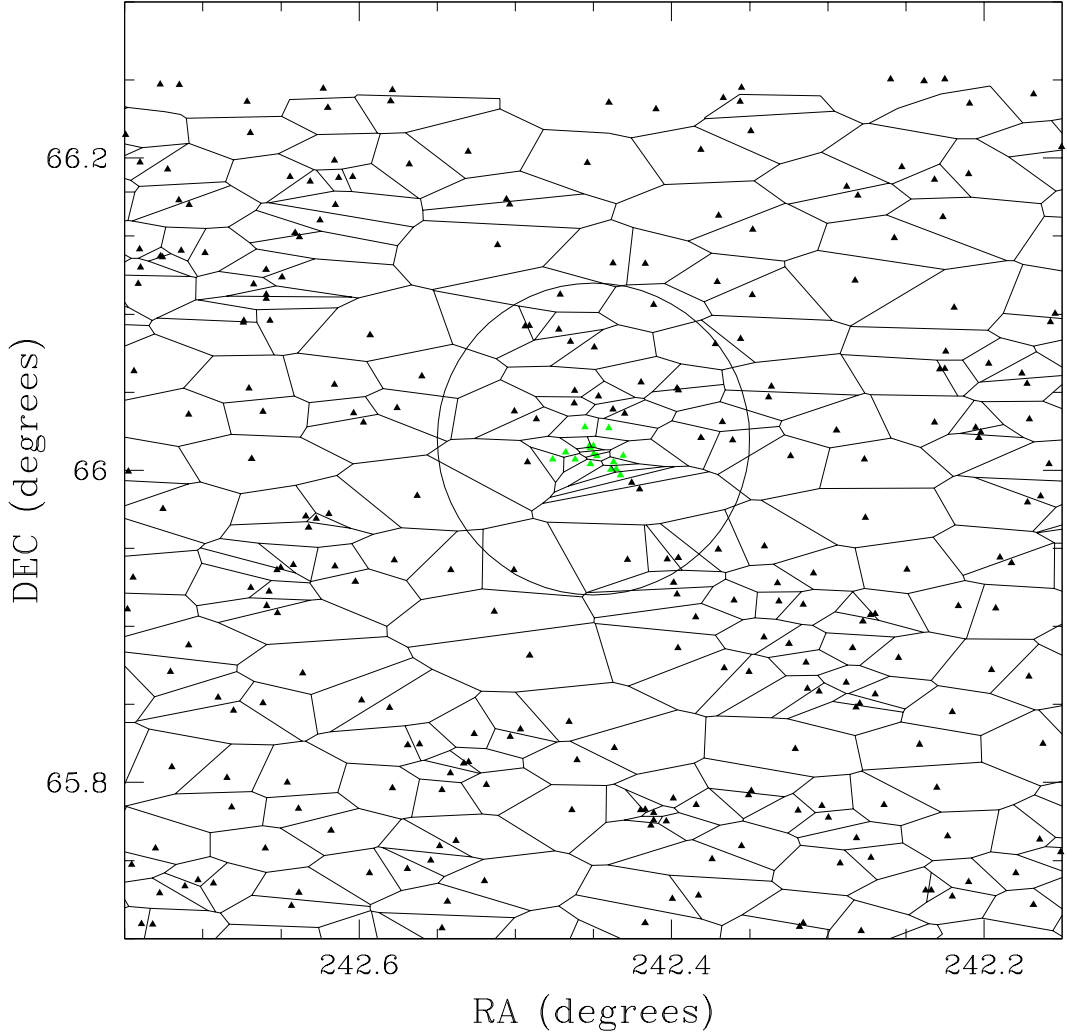}
\caption{Voronoi tessellation on the galaxy distribution for ObsID 2127. 
Only galaxies satisfying the color cut expected for a cluster red-sequence 
at a redshift of 0.475 are depicted. The area enclosed within the circle is 
a previously unknown cluster at an estimated VTP redshift of $0.527$. This 
cluster was also detected as an extended X-ray source and is listed as 
object CXOMP~J160948.4+660057 in our data tables.}
\label{VTPcells}
\end{figure}

The VTP algorithm partitions the galaxy spatial plane into polyhedral 
cells, each containing a single unique galaxy (Voronoi cell). The cell 
size is determined by the distance between nearest neighbors and encloses 
the maximum area nearest to a given galaxy (see Figure~\ref{VTPcells}). 
Galaxy clusters are detected as overdensities in the number of Voronoi 
cells (grouped together using a percolation technique) per unit area 
\citep{Ramella01}. A slightly modified approach is to use the 
inverse of the area contained within each Voronoi cell 
\citep{Ebeling93,Kim02}. The significance of each galaxy overdensity is 
computed by comparing the density distribution of the galaxy catalog with 
that expected for a random distribution of Voronoi cells --- the so-called 
Kiang distribution \citep{Kiang66}. Overdense regions composed of adjacent 
Voronoi cells are flagged as potential clusters if their density is greater 
than a specified threshold. A random fluctuation in the background can 
potentially exceed the imposed threshold constraint and thus be counted 
as a real cluster. This contamination is minimized by computing the 
probability (based on simulations) that a given detection is a random 
background fluctuation, and then only including regions above an acceptable 
level (see \S 3.2.2).

To improve the contrast of cluster galaxies with respect to the background 
field population, we have implemented a refined version of the VTP method 
that takes advantage of the existence of the red-sequence in the 
color-magnitude relation of early-type cluster galaxies (e.g., Baum 1959; 
Sandage \& Visvanathan 1978; L\'{o}pez-Cruz, Barkhouse, \& Yee 2004). The 
red-sequence for early-type cluster galaxies shifts to progressively redder 
observed colors as the 4000\AA\, break moves through the filter bandpasses 
with increasing redshift (i.e., the $K$-correction; Humason, Mayall, 
\& Sandage 1956; Oke \& Sandage 1968). The position of the cluster 
red-sequence in the color-magnitude plane can be used as an estimator of 
redshift \citep{Gladd00,Lopez04}. Thus, choosing appropriate filters 
will enable foreground and background galaxies to be culled to minimize 
the contamination from the field galaxy population. As an example, 
\citet{Kim02} applied the VTP algorithm to SDSS galaxy catalogs constructed 
by selecting galaxies relative to the expected red-sequence in the 
$g^{\ast}-r^{\ast}$ vs. $r^{\ast}$ color-magnitude plane for clusters at 
various redshifts. The color width blueward of the red-sequence for each 
redshift slice was chosen to be relatively broad 
($\Delta(g^{\ast}-r^{\ast})\sim 0.6$ mag; see their Figure 2). 

\begin{figure}
\figurenum{4}
\plotone{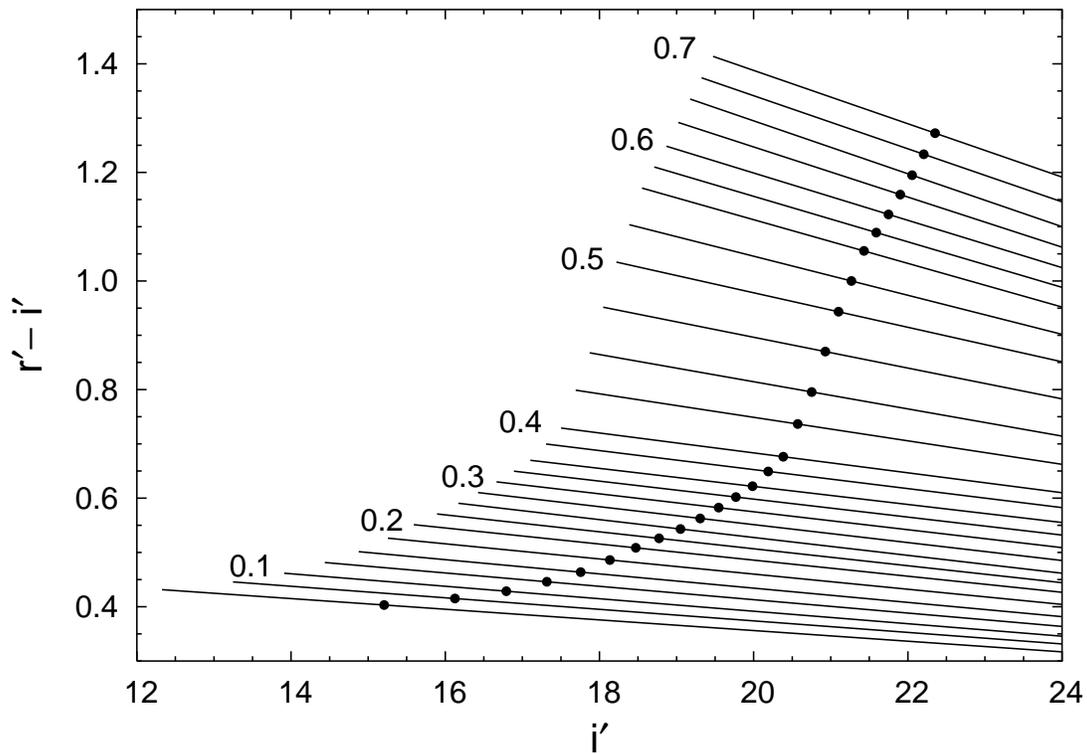}
\caption{Red-sequence model color slices used to select galaxies at various 
redshifts (indicated to the left of the lines) as part of the cluster 
detection process using VTP. Galaxy catalogs are constructed for each 
red-sequence by selecting galaxies with $r^{\prime}-i^{\prime}$ color within 
$\pm 0.1$ mag of the red-sequence line. Overlapping color slices allow 
us to completely sample the color-magnitude plane for $z=0.05-0.70$. The 
solid points depict the $i^{\prime}$-band magnitude of $m^{\ast}$ for our 
sampled redshift range.} 
\label{VTPslices}
\end{figure}

The ChaMP optical data consists of magnitudes measured in the $g^{\prime}$-, 
$r^{\prime}$-, and $i^{\prime}$-band. Since we are interested in assembling 
a cluster sample that extends to high redshift ($z> 0.5$), we have elected 
to use the $r^{\prime}-i^{\prime}$ color to select galaxies since 
$g^{\prime}-r^{\prime}$ becomes degenerate at $z\lesssim 0.4$ (T. Kodama 
2004, private communications). The $r^{\prime}-i^{\prime}$ color will 
allows us to sample cluster red-sequences out to $z\sim 0.7$ 
(see Figure~\ref{VTPslices}). The basic procedure is to construct catalogs 
containing galaxies with a $r^{\prime}-i^{\prime}$ color distribution that 
matches a particular red-sequence for a given redshift. Catalogs are produced 
for red-sequences that sample the redshift range from $z=0.05-0.70$. VTP is 
then applied to each galaxy catalog and the most significant detections 
(as flagged by VTP) are included in the final cluster compilation for a 
given field. The advantage of this technique over the standard VTP method 
\citep[e.g.,][]{Ramella01} is that the ``noise'' from field galaxies is 
reduced, and also that the redshift of the detected cluster can be 
estimated from the catalog yielding the greatest detection significance. 
Several recent studies have been successful in using color-cuts relative 
to the red-sequence to search for clusters using optical data 
\citep[e.g.,][]{Gladd00,Goto02,Nichol04,Hsieh05}.
 
For the location of the red-sequence in the color-magnitude plane, we 
adopt the models of \citet{Kodama97} transformed to the SDSS filters 
(T. Kodama 2004, private communications). Each galaxy catalog is 
generated for a specific red-sequence by selecting galaxies with a 
$r^{\prime}-i^{\prime}$ color within $\pm 0.1$ mag of the red-sequence line 
(all galaxies are corrected for galactic extinction prior to the selection 
process; see Table 2). We choose a color width on either side of the 
red-sequence of $0.1$ mag since the measured dispersion of cluster galaxies 
along the red-sequence is $\sim 0.07$ mag 
\citep[e.g.,][]{Bower92,Ellis97,Lopez04}. To insure that our galaxy catalogs 
sample the complete range in color for our expected cluster redshift 
distribution, we construct galaxy samples for 27 red-sequences from 
$z=0.05-0.70$ (Figure~\ref{VTPslices}). The density of these model 
red-sequences in the 
color-magnitude plane allows our galaxy color slices to overlap with 
adjacent regions by $\sim 0.1$ mag in $r^{\prime}-i^{\prime}$. Thus we 
completely cover the color-magnitude plane with overlapping color slices 
in the region expected for our cluster redshift range. The overlapping 
color slices prevent us from missing clusters whose red-sequence may fall 
between our adopted 27 red-sequence (i.e., redshift) models.  

In addition to selecting galaxies by color, we also restrict the 
magnitude range for each color slice to enhance the cluster signature 
above the background. We have elected to include galaxies with 
$i^{\prime}$-band magnitudes in the range of $m^{\ast}_{i^{\prime}}-3$ to 
$m^{\ast}_{i^{\prime}}+4$, where $m^{\ast}_{i^{\prime}}$ is the apparent 
magnitude 
of the turnover in the Schechter function representation of the 
cluster luminosity function \citep{Schechter76}. The $m^{\ast}_{i^{\prime}}$ 
value was derived by taking the value of 
$M^{\ast}_{R_{c}}=-22.20+5\,\mbox{log}\,h_{50}$ from \citet{Barkhouse03} for 
a sample of 57 low-redshift clusters and transforming to 
$M^{\ast}_{i^{\prime}}=-21.52+5\,\mbox{log}\,h_{70}$ using the equations of 
\citet{Frei94}. For each redshift interval, we transformed 
$M^{\ast}_{i^{\prime}}$ 
into $m^{\ast}_{i^{\prime}}$ using the appropriate luminosity distance, and 
$K$-correction tabulated in \citet{Frei94} for an early-type population 
(no correction for galaxy evolution was applied). 
We note that galaxies can be included using a smaller magnitude range 
if CCD image saturation occurs at a magnitude fainter than 
$m^{\ast}_{i^{\prime}}-3$ 
or if the magnitude limit (measured as the turnover in the galaxy counts) 
is brighter than $m^{\ast}_{i^{\prime}}+4$ (see Table 2 for the turnover 
magnitude limits). Our magnitude restriction thus insures that galaxies 
are included 
in a consistent fashion for each red-sequence (i.e., redshift) slice while 
excluding bias due to saturation and incompleteness effects.  

The final culling that we conducted on our galaxy catalogs was to 
exclude all objects with a stellarity class $> 0.9$ in the 
$i^{\prime}$-band (i.e, ``star-like'' objects), as output by 
SExtractor \citep{Bertin96}. The median seeing of our optical images 
is $1.0\arcsec$, which corresponds to a linear distance of $\sim 7$ kpc 
at $z=0.7$ for our adopted cosmology. The stellarity cut thus enabled us 
to produce a statistically high fidelity sample of galaxies, facilitating 
red-sequence detection.

Once galaxy catalogs for our 27 red-sequence slices had been compiled for 
each field, we ran VTP independently on our 56 mosaic fields. For this paper, 
we restrict the analysis and discussions to a sample of 36 mosaic fields 
(see Table 2) that contain VTP-detected optical cluster candidates that are 
located within the field-of-view of the {\it Chandra} ObsIDs (total sky 
coverage of 6.1 square degrees). This allows us to fairly compare cluster 
detection fractions, and measure optical and X-ray properties of matched, 
serendipitously-detected, extended sources. 

\subsubsection{Monte Carlo Simulations}

In addition to supplying galaxy catalogs, the detection threshold and 
rejection probability limit against random background fluctuations 
must be set as inputs to VTP. In \citet{Ramella01} the detection threshold 
was set at the 80\% confidence level and detected sources with a probabilily 
$> 5\%$ of being random background fluctuations were rejected from the 
final cluster compilation.

To determine the best choice of detection and rejection parameters, we 
conducted a series of extensive Monte Carlo simulations. We adopted the 
contamination rate as a benchmark to compare different runs of VTP. The 
contamination rate is defined as $\mbox{C}=\mbox{N}_{S}/\mbox{N}_{D}$, 
where $\mbox{N}_{S}$ is the number of detected clusters from a simulated 
galaxy catalog, and $\mbox{N}_{D}$ is the number of detected clusters based 
on the original catalog using the same VTP input parameters as $\mbox{N}_{S}$. 
The goal of the simulations is to run VTP for a range in values of the 
threshold and rejection probability to determine which values minimize 
the contamination rate while maximizing the detection of real clusters.

The simulated cluster catalogs were constructed using three different 
procedures to alter the original galaxy catalog generated by SExtractor 
for each mosaic field: randomizing positions, and shuffling magnitudes 
and colors. For the first procedure, we randomized galaxy positions while 
keeping the magnitudes and colors the same. Next, 1000 simulated galaxy 
catalogs were constructed for each set of threshold and rejection probability 
pairs (30 combinations in total were used). Each of the 1000 simulated 
catalogs were divided into 27 red-sequence slices (as described in 
\S 3.2.1) and VTP run separately on each. The average contamination 
rate, $\mbox{C}_{ave}$, was calculated for each set of 1000 simulated 
catalogs for each ordered pair of the detection threshold and random 
probability values. In total, VTP was run on 30,000 simulated galaxy 
fields, each containing 27 red-sequence slices (i.e., 810,000 executions 
of the VTP algorithm). The minimum value of the contamination rate was 
found to be $\mbox{C}_{ave}=12.8\%$ for a detection threshold value 
of 90\% confidence level and a random probability limit of 5\% (i.e., 
overdensities having a probability $> 5\%$ of being a random background 
fluctuation are rejected). Note that these values are very similar to those 
used in \citet{Ramella01}.

The second procedure we used to measure the false-positive rate involved 
shuffling (without replacement) the $i^{\prime}$-band magnitudes while 
maintaining the original galaxy colors and positions. The magnitudes 
are shuffled rather than assigned randomly to preserve the galaxy 
brightness distribution as measured for the real data. The simulations 
were undertaken in the same manner as conducted previously for the random 
galaxy position catalogs. The minimum contamination rate for the shuffled 
magnitude catalogs was determined to be $\mbox{C}_{ave}=20.76\%$. This 
minimum value coincided with the same pair of VTP detection parameters as 
that found above using the randomized position catalogs (i.e., 90\% 
confidence level and a random probability threshold of 5\%).

For the final test, galaxy colors were shuffled while retaining the 
original magnitudes and positions. Using the same number of simulations 
as described above, the minimum contamination rate was found to be 
$\mbox{C}_{ave}=7.52\%$. This minimum value coincidently occurred for the 
same pair of VTP detection parameters as found for the previous two 
independent sets of simulations. Thus, our Monte Carlo simulations 
demonstrate that executing VTP with a confidence level of $90\%$ and 
a random probability limit of $5\%$ will minimize the contamination rate. 

The false-positive tests show that shuffling the galaxy colors produces the 
smallest contamination rate, while shuffling the $i^{\prime}$-band magnitudes 
exhibits the largest. This is directly related to the red-sequence slices 
that we employ prior to running the VTP detection. Since cluster 
red-sequences based on $r^{\prime}-i^{\prime}$ vs. $i^{\prime}$ 
are approximately 
horizontal for a wide range in redshift (see Figure~\ref{VTPslices}), 
shuffling galaxy $i^{\prime}$ magnitudes, while maintaining their original 
color, has the least effect on washing out the signature of real clusters 
in the simulated catalogs. Conversely, shuffling colors is expected to have 
the greatest impact since the red-sequence of real clusters will be smoothed 
out. Although our simulations were not designed to reproduce the two-point 
correlation function for field galaxies (which would be expected to have 
a higher contamination rate than a randomized position catalog; Gilbank 
et al. 2004), the simulations produced by shuffling the colors should be 
a reasonable estimate of the contamination rate. Thus, we surmise that our 
red-sequence VTP method suffers from a contamination rate of $< 20\%$.

It is important to note that our simulations are not designed to 
provide a perfect measure of the false-positive rate, but rather serve 
as an indicator on how to tune our cluster detection code. We therefore 
adopt a detection threshold of a 90\% confidence level and a cutoff 
probability for a random fluctuation of 5\% for our red-sequence VTP 
application.  

\subsubsection{Final Optically-Selected Cluster Sample}

Once the VTP cluster detection was implemented for the 27 red-sequence 
slices per field with the input parameters derived from our Monte Carlo 
simulations, we merged our candidate clusters into a single catalog. The 
redshift assigned to each cluster is determined by selecting the red 
sequence slice that maximized the product of the confidence level and 
contrast above background. These parameters are output by the VTP algorithm 
and are {\it not} the same as the confidence threshold and random 
fluctuation level set as input to VTP \citep{Ramella01}. As an 
additional step to minimize spurious sources, we include only those 
detections that have an output-measured confidence level $>99\%$ (this 
value was selected by visually inspecting the VTP source catalog). This 
culling procedure was also implemented for the simulations described in 
\S 3.2.2. For cluster centroids that are $<2\arcmin$ apart and separated by 
$\Delta z<0.1$ (estimated from the specific red-sequence slice from 
which the cluster was detected with maximum probability), we merged the 
candidates into a single cluster. Finally, we visually inspected our 
cluster candidates to exclude spurious or contaminated clusters/groups 
from the final sample. 

\section{Results}

With the construction of two independent cluster compilations derived from 
the same sky area, we are able to compare and contrast the 
attributes of our two samples, each selected from different wavelength 
regimes. In this section, we will describe the properties of each cluster 
sample and compare cluster detection techniques. 

\subsection{Optical Cluster Properties}

Our final sample of optical cluster candidates selected using our red-sequence 
VTP technique --- independent of X-ray detections --- contains 115 sources 
measured from 36 mosaic optical fields. As mentioned previously, this sample 
contains only optical VTP detections from regions that overlap with the 
{\it Chandra} sky coverage, and excludes {\it Chandra} PI targets (of which 
15 clusters are detected by our VTP method). 

\begin{figure}
\figurenum{5}
\plotone{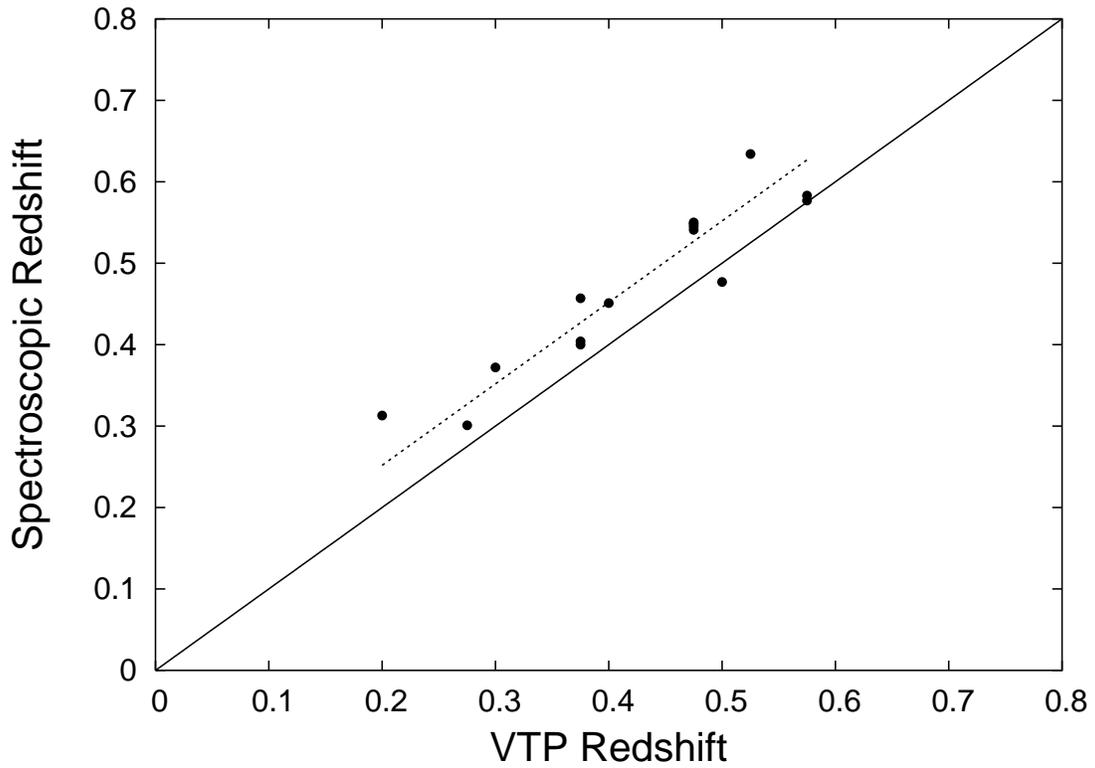}
\caption{Comparison between spectroscopic and VTP derived redshifts for a 
sample of 15 previously known rich clusters ($0.3\leq z\leq 0.7$) within our
sampled fields. The dashed line 
represents a fit to the offset between the two redshift measurements 
($\Delta z=+0.052$).}
\label{Spec-VTPredshifts}
\end{figure}

A main benefit of using VTP on galaxies selected relative to 
the red-sequence is that we are able to assign a photometric 
redshift to each cluster candidate. Since a perfect match between the 
filter transmission function used to obtain our mosaic data and that 
assumed for the red-sequence models is {\it not} expected, estimated 
red-sequence redshifts may be systematically offset from the ``true'' 
values. To quantify this effect, we compared the photometric redshifts 
estimated from VTP with spectroscopically determined redshifts for a 
sample of 15 known clusters ($0.3\leq z\leq 0.7$) that are contained 
within our ChaMP mosaic fields. These clusters are not included in our 
final cluster sample since they are either {\it Chandra} PI targets or 
not found within the spatial coverage of our 36 optical ChaMP fields that 
overlap our X-ray sky coverage. In Figure~\ref{Spec-VTPredshifts} we 
plot spectroscopic redshifts versus VTP redshifts for the 15 cluster 
sample, which illustrates a systematic offset in the VTP redshifts 
relative to the spectroscopic values in the sense that the photometrically 
derived cluster redshifts are underestimated. A line of slope unity yields 
a good fit to the data with an offset of $\Delta z=+0.052$, and a 
dispersion of $0.033$ (dashed line in Figure~\ref{Spec-VTPredshifts}). To 
improve VTP redshift estimates for our cluster sample, we add the correction 
term of $\Delta z=0.052$ to all photometrically derived redshifts. For X-ray 
detected clusters, the VTP redshift is included in Table 3. Table 4 
lists the cluster candidates detected solely from the application of 
the red-sequence filtered VTP on the 36 mosaic fields that overlap 
our {\it Chandra} fields tabulated in Table 1. Corrected VTP redshifts 
and limits to $f_X$ and $L_X$ are also provided there.

\begin{figure}
\figurenum{6}
\plotone{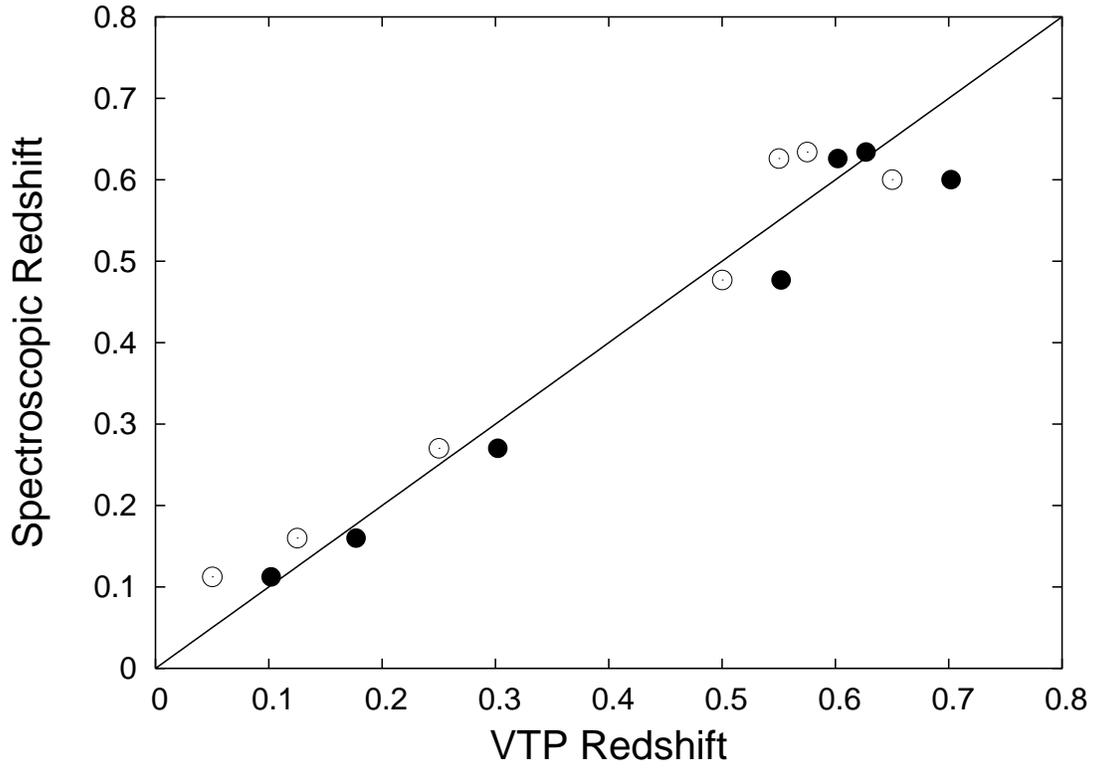}
\caption{The solid points are VTP redshifts with correction applied and open 
circles are VTP redshifts without correction. The application of the 
correction lowers the dispersion of the VTP-estimated redshifts from the 
corresponding spectroscopic measurements (solid line) from 0.05 to 0.03.}
\label{VTP-correct}
\end{figure}

In Figure~\ref{VTP-correct} we present a comparison between 7 
ChaMP serendipitously-detected extended X-ray sources that are matched 
in the optical and have both spectroscopic and VTP estimated redshifts. 
The open circles are for VTP derived redshifts without the $\Delta z$ 
correction applied, while the solid circles include the correction. 
The application of the redshift correction lowers the dispersion of the 
VTP-estimated redshifts from the corresponding spectroscopic measurements 
from 0.05 to 0.03.

\subsection{X-ray Cluster Properties}

\begin{figure}
\figurenum{7}
\plotone{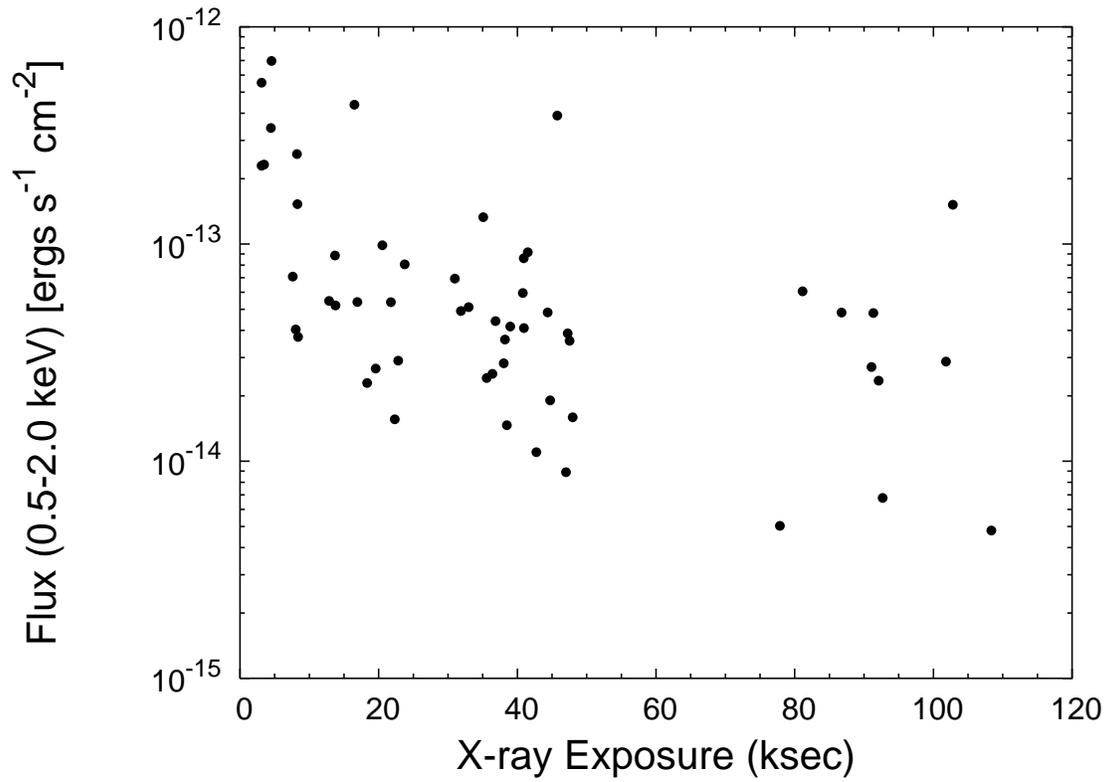}
\caption{X-ray flux (0.5-2 keV) of extended X-ray sources as a 
function of vignetting-corrected exposure time. Approximately 82\% of the 
extended sources are detected from fields with exposure times $< 50$ ksec.}
\label{flux-exp}
\end{figure}

The availability of multi-wavelength archival data affords the opportunity 
to conduct a survey for specific objects with minimal investment in 
observing time. In Figure~\ref{flux-exp} we show the distribution of 
X-ray flux in the 0.5-2 keV energy band for our 55 extended sources 
as a function of vignetting-corrected exposure time. The median flux of 
the sample is 
$f_X=(4.84\pm 0.15)\times 10^{-14}~\mbox{erg}~\mbox{s}^{-1}~\mbox{cm}^{-2}$ 
with $\sim 82\%$ of the sources detected in exposures $<50$ ksec. 

\begin{figure}
\figurenum{8}
\plotone{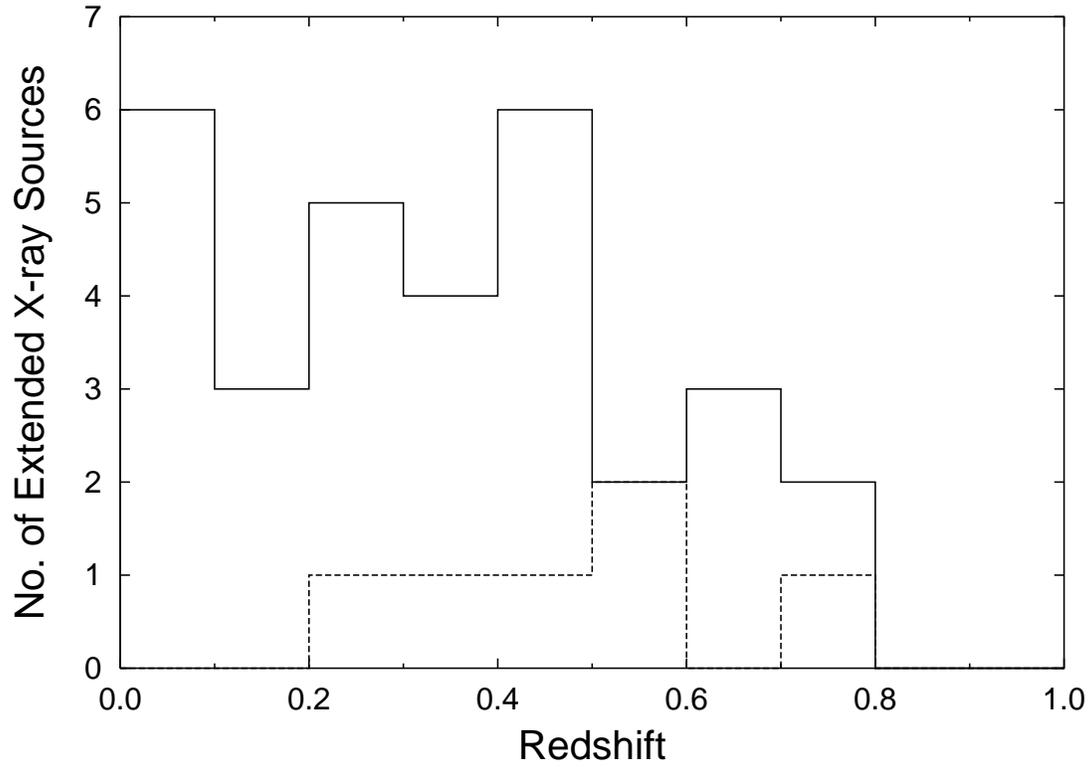}
\caption{The redshift distribution of 31 extended X-ray sources with 
either spectroscopic- (solid line) or VTP-estimated redshifts 
(dashed line). Six sources are associated with low-redshift single 
galaxies ($z<0.1$). The redshift distribution for the cluster-only sample 
peaks at $z\sim 0.4$}
\label{redshift-his}
\end{figure}

A histogram of the redshift distribution for 31 of our X-ray sources with 
redshifts is presented in Figure~\ref{redshift-his}. Redshift values are 
measured either spectroscopically from ChaMP (or other published sources; 
see Table~5) or from our red-sequence filtered VTP technique. Six sources 
at $z< 0.1$ are coincident with nearby single galaxies, as ascertained 
by examining their positions in our mosaic data \citep{Kim06}. The final 
extended X-ray source catalog thus contains 49 
clusters and 6 low-$z$ galaxies. The average redshift of the cluster-only 
sample is $\overline{z}=0.41$, which corresponds to the peak in the redshift 
histogram distribution. 

The distribution of the X-ray luminosity (0.5-2 keV) for our 31 sources 
with estimated redshifts is depicted in Figure~\ref{luminosity-his}. The 
X-ray luminosity spans the range from rich clusters 
($L_X\sim 10^{43}-10^{45}~\mbox{ergs}~\mbox{s}^{-1}$; Rosati et al. 2002) 
to poor groups ($L_X\sim 10^{41}-10^{43}~\mbox{ergs}~\mbox{s}^{-1}$; 
Mulchaey 2000), and bright starburst galaxies 
($L_X\gtrsim 10^{39}~\mbox{ergs}~\mbox{s}^{-1}$; Fabbiano 1989). The median 
value of the luminosity for the complete sample of extended sources is 
$L_X=(2.18\pm 0.18)\times 10^{43}~\mbox{ergs}~\mbox{s}^{-1}$, which 
increases to $L_X=(2.65\pm 0.19)\times 10^{43}~\mbox{ergs}~\mbox{s}^{-1}$ 
when the six low-$z$ galaxies are excluded. 

In \S 4.3 we compare the cluster candidates detected by our red-sequence 
VTP method with the extended X-rays sources in the sky area where our
X-ray and optical imaging overlaps (6.1 square degrees). In several 
instances, the X-ray counterpart to the optical detection is absent. 
For these cases, we compute the upper limit X-ray flux by measuring photon 
counts in a $84\arcsec$ aperture radius centered on the optical cluster 
position as 
defined by the location of the brightest cluster galaxy (BCG) or the centroid 
position output by the VTP algorithm. The $3\sigma$ upper limits are computed 
by extrapolating the counts to infinity using the $\beta$-model (see \S 3.1) 
and then converting to luminosity limits using the redshift estimated by the 
VTP process, or if available, that derived spectroscopically from ChaMP or 
other published sources. The $84\arcsec$ radius represents a factor of 
$4\times$ the median core radius of the 55 extended X-ray sources and is 
equivalent to a core radius of $460~\mbox{kpc}$ at $z=0.41$ (median redshift 
of the cluster-only sample) for our adopted cosmology. The X-ray flux and 
luminosity upper limits, for the 0.5-2 keV energy band, are tabulated in 
Table 4 for the 102 clusters with optical-only detections.

In Figure~\ref{luminosity-his} the X-ray luminosity distribution of the 
upper limits (dashed line) is compared to the distribution of the detected 
X-ray sources. From this figure it appears that many of the optical sources 
not detected in the X-rays have luminosities characteristic of groups and 
clusters (median $L_X\sim 10^{43}~\mbox{ergs}~\mbox{s}^{-1}$), thus 
making them harder to detect in the shallower X-ray fields at the expected 
faint flux levels based on 
the $L_X-T_X$ and $\mbox{M}_{tot}-T_X$ relations \citep[e.g.,][]{Ettori04}. 
The distribution of X-ray luminosity with redshift is depicted in 
Figure~\ref{luminosity-redshift}. The circles represent the sample of 
25 X-ray-detected clusters with estimated redshifts while the triangles 
depict the six extended X-rays sources identified with low-$z$ single 
galaxies. Also plotted are the X-ray luminosity upper limits (arrows) for 
sources detected only in the optical by VTP. Sources with spectroscopic 
redshifts are shown as the larger symbols while objects with only VTP 
estimated redshifts are marked with the smaller symbols. Even though the 
X-ray flux limit varies from field-to-field due to the wide range in exposure 
times (see Figure~\ref{flux-exp}), the solid line in 
Figure~\ref{luminosity-redshift} indicates the luminosity for a flux limit of 
$f_X~\mbox{(0.5-2 keV)}=1.5\times 10^{-14}~\mbox{ergs}~\mbox{s}^{-1}
~\mbox{cm}^{-2}$. This flux limit is plotted for comparison purposes only, 
but it is a reasonable estimate of the overall flux limit of the survey. 

To compare the ChaMP sources with previous surveys, we plot in 
Figure~\ref{luminosity-redshift} the 447 clusters (``+'' symbols) from the 
{\it ROSAT}-ESO Flux-Limited X-Ray (REFLEX) Galaxy Cluster Survey 
\citep{Bohringer04}. The luminosity range probed by our sample is at least 
an order of magnitude below that of previous large-area ROSAT 
samples like REFLEX. This provides newly extended coverage of the 
$L_X$-redshift plane down to group luminosities even at significant 
($\sim$0.5/$\mbox{H}_{\mbox{o}}$) look-back times, thus paving the way 
for studies of cluster evolution that can take into account not just 
cosmic time but also mass/luminosity/temperature effects. In addition to 
the REFLEX cluster survey, we also plot 200 galaxy clusters (``x'') from 
the 160 Square Degree {\it ROSAT} Survey \citep{Mullis03}. This cluster 
sample is well-matched to our ChaMP sample in terms of the measured range in 
luminosity and redshift. The median redshift of the 160 Square Degree 
Survey clusters is $z=0.25$, which is lower than the median redshift of the 
ChaMP cluster-only sample ($z=0.41$). 

The core radius for each extended X-ray source is measured with 
$\beta=0.67$ due to the small number of photons expected for the 
majority of our sources. A fit to the surface brightness profile 
with $\beta$ fixed at 0.67 yields a core radius accurate to 
$\pm 20\%$ for $0.6<\beta <0.8$ \citep{Jones99}. The X-ray 
surface brightness radial profile for a typical sample of 
nine extended sources is given in Figure~\ref{SB-radial}. The 
best-fit $\beta$-model (solid line) along with the associated $1\sigma$ 
uncertainty (dashed lines) are presented. In Figure~\ref{coreradius-his} 
a histogram of the core radius distribution 
is presented for our 55 extended sources. The median angular core radius 
($r_{c}=21.37\pm 2.22\arcsec$) is depicted by the dashed line. If 
the 6 low-$z$ galaxies are excluded, the median core radius increases 
to $r_{c}=22.08\pm 1.67\arcsec$. This value is comparable to the 
typical core radius for extended sources detected using 
the ROSAT HRI \citep{Moretti04} over a similar flux and redshift range. 

To search for possible evolution in the measured core radius 
as a function of redshift, we plot in Figure~\ref{coreradius-redshift} the 
angular core radius (arcsec) and the metric core radius 
(kpc) as a function of redshift. The top panel reveals 
that the detectability of clusters of different observed core radii is not 
a strong function of redshift. When we plot the core radii in linear 
units (bottom plot), no direct correlation with redshift is observed for our 
sample. This result is in agreement with the study of \citet{Vikhlinin98a} 
which found no obvious evolution of the core radius with redshift for 
a sample of 203 clusters from the 160 Square Degree {\it ROSAT} Survey. 
However, we do see an increase in the dispersion of core radii with redshift 
in the bottom panel of Figure~\ref{coreradius-redshift}. This may be a 
volume effect --- more luminous clusters with larger core radii are rare, 
and therefore only well-sampled at higher redshifts.

The core radius versus X-ray luminosity is plotted in 
Figure~\ref{coreradius-flux} for the 31 extended X-ray sources with 
either spectroscopic or VTP estimated redshifts. A correlation exists 
between the core radius and $L_X$ in the sense that the more X-ray 
luminous sources are physically more extended. A correlation between 
the core radius and X-ray luminosity is not unexpected given that the 
physical size of X-ray clusters has been found to increase 
with X-ray temperature, and hence luminosity \citep{Mohr00}. This effect was 
also seen by \citet{Jones99} for a heterogeneous sample of 368 low-redshift 
($z<0.2$) X-ray clusters imaged by {\it Einstein}. To determine which of the 
variables; redshift, $L_X$ or $f_X$ presents the primary correlation with 
core radius $r_{c}$, we apply partial correlation analysis to the 31 
serendipitous X-ray extended source detections with redshift measurements 
(from either spectroscopy or VTP redshift estimates). For our sample of 
$\mbox{N}=31$, the correlation of $r_{c}$ with $L_X$ remains highly 
significant when holding either $f_X$ or redshift constant. While, as 
expected, our effective flux limit creates the strongest simple 
correlation between $L_X$ and redshift, of all the other combinations 
tested, $r_{c}$ depends most strongly on $L_X$ (partial 
Kendalls $\tau= 0.439$ with $\sigma=0.116$; \citealt{akritas96}). 
A power-law fit to the X-ray luminosity--redshift data yields 
$r_c\propto L_X^{0.48\pm 0.04}$ and is displayed as the 
solid line in Figure~\ref{coreradius-flux}.    

Galaxy mergers are expected to have a greater impact on galaxy evolution 
in the group environment rather than at the center of rich clusters due to 
the lower velocity dispersion of the group members \citep[e.g.,][]{Dub98}. 
Simulations predict that the end result of galaxy mergers in groups will 
probably be the formation of a single elliptical galaxy with an extended 
X-ray halo \citep{Mamon87,Barnes89}. These ``fossil groups'' provide 
important information on the evolution of galaxies and the ICM in these 
type of locales \citep{Ponman94,Vikhlinin99,Jones00,Ulmer05}. An expected 
signature of fossil groups is the presence of extended X-ray emission 
centered on a luminous early-type galaxy. To determine if our six single 
galaxies with extended X-ray emission are fossil groups, we compared the 
spatial extent of the X-ray and optical emission. For all six galaxies, 
the X-ray emission is more concentrated than the optical light with no 
evidence to suggest that some fraction of the X-ray emission is due 
to a group ICM. We thus conclude that none of our six galaxies are 
associated with a fossil group.

\subsection{Comparison of X-ray and Optical Clusters}

We cross-correlated the extended X-ray source catalog, compiled from 
the wavelet decomposition technique, with the optical cluster sample 
constructed from the red-sequence filtered VTP method. From the sample 
of 55 extended X-ray sources, 6 were found to be associated with 
low-redshift galaxies. Of the 49 X-ray extended sources not associated with 
low-redshift galaxies, only 28 are located in X-ray fields that overlap 
with our MOSAIC pointings (6.1 sq. deg.; see Table 2). The fraction of 
matches between the X-ray and optical cluster catalogs is 46\% (13 out 
of 28), with a median redshift of $z_{med}=0.477\pm 0.202$, where the 
uncertainty is the rms of the dispersion. In Table 5, we tabulate source 
information from the literature for objects near the position of each 
extended X-ray source as archived by the NASA/IPAC Extragalactic 
Database.\footnote{This research has made use of the NASA/IPAC 
Extragalactic Database (NED) which is operated by the Jet Propulsion 
Laboratory, California Institute of Technology, under contract with the 
National Aeronautics and Space Administration.}

A $46\%$ match of the X-ray clusters to our optical VTP sources may 
not seem surprising given that the variation in the X-ray 
and optical exposure times (see Tables 1 and 2) affects the limiting 
flux reached in either passband. To test whether optical magnitude 
limits have a direct impact on the matched rate for our sample, the 
turnover magnitude\footnote{The turnover magnitude is the magnitude 
at which the differential galaxy counts begin to decrease with increasing 
magnitude due to incompleteness.} for fields containing matched clusters 
is compared to fields without matched sources (see Table 2 for the 
turnover magnitudes for each field). The turnover magnitude for the 
fields with optical-X-ray matches varies from $i^{\prime}=23.12$ mag 
to $i^{\prime}=24.62$ mag, with a median value of $23.62 \pm 0.41$. For 
the optical fields containing no detected extended X-ray sources, the 
turnover magnitude varies from $i^{\prime}=22.12$ mag to 
$i^{\prime}=24.62$ mag, with a median value of $23.62 \pm 0.78$. The 
equality of the median turnover magnitudes for the optical fields 
containing detected and non-detected X-ray sources (most of the fields 
are identical between the two samples) suggests that we are not missing 
a large fraction of our extended X-ray sources in the optical due to a 
variation in the optical magnitude limit between the two samples.

A possible explanation for the 46\% matched-rate between our X-ray clusters 
and optical VTP counterparts is that the X-ray flux limit is much fainter 
relative to the optical limit for fields which lack optical matches than 
for fields containing matched sources. However, a comparison of our 
measured flux limit ratios indicates that the difference between the 
matched and unmatched samples are statistically insignificant. Indeed, 
the ChaMP optical follow-up was originally designed to probe similar 
$f_X/f_o$ populations in each field by tuning optical magnitude 
limits to the X-ray exposure times for each field \citep{Green04}. 

In Figure~\ref{iMag-Flux} we plot the $i^{\prime}$-band magnitude 
optical field limit versus X-ray flux for the 28 extended X-ray sources 
(excluding the low-$z$ galaxies) that overlap our optical fields. From 
this figure we see that for optical fields with turnover magnitude limits 
brighter than $i^{\prime}=23$, all four extended X-ray sources are not 
detected in the optical data by VTP (optical data is too shallow to allow a 
robust detection). The area of the plot having X-ray sources with 
bright X-ray fluxes and faint optical field limits contains several sources 
which are undetected by VTP. If we restrict ourselves to fields with 
optical magnitude limits fainter than $i^{\prime}=23$ and 
$f_{X}>10^{-14}~\mbox{ergs}~\mbox{s}^{-1}~\mbox{cm}^{-2}$, we find 10 
X-ray sources without optical matches. A detailed visual inspection of   
our optical images coincident with these 10 sources shows that 2 are 
most-likely undetected by VTP due to the presence of a nearby bright 
star, 4 sources have very faint galaxies near the X-ray 
centroid but too faint to be flagged by VTP, and the remaining 4 show no 
conclusive evidence of being clusters in the optical (i.e., the area looks 
``field-like'' in nature). These undetected extended X-ray sources are 
prime candidates for deeper optical/near-IR follow-up imaging.

Our VTP red-sequence detection may be missing optical clusters with 
a large fraction of blue galaxies. In addition, there is the possibility 
that some of the unmatched X-ray sources are high-$z$ clusters whose 
galaxies are too faint in these optical filters to be detected by our 
VTP algorithm. Of the 15 extended X-ray sources without VTP optical matches, 
only three have measured spectroscopic redshifts ($z=0.3768$, $0.4118$, 
and $0.4256$), with X-ray luminosities ranging from 
$(4.4-26)\times 10^{42}~\mbox{ergs}~\mbox{s}^{-1}$. These three extended X-ray 
sources have X-ray point sources embedded within them and were targets 
of the ChaMP spectroscopic AGN follow-up program. These sources were not 
detected by VTP since only a couple galaxies were observed within the 
red-sequence slice consistent with the redshift of the extended X-ray 
source (2 galaxies for the $z=0.3768$ source, 3 galaxies for the 
$z=0.4118$ source, and 1 galaxy for the $z=0.4256$ source). To test 
whether the remaining 12 sources have properties consistent with our 
X-ray/optically matched sample, we include these objects on a plot of 
$L_{X}$ versus redshift (see Figure~\ref{luminosity-redshift3}), which 
is similar to Figure~\ref{luminosity-redshift}. To estimate a redshift, 
we simply use the extinction-corrected $i^{\prime}$ magnitude of the galaxy 
located closest to the X-ray centroid and assume that it is an 
$M^{\ast}$ elliptical galaxy. We apply K- and evolution-corrections based 
on a \citet{Bruzual03} passive evolution model for early-type galaxies 
with a formation redshift of $z=3$. Figure~\ref{luminosity-redshift3} 
shows that 5 of the 12 sources would have properties consistent with $z>0.8$ 
clusters and thus could be at redshifts greater than our 
X-ray/optically matched sample. If we assume the galaxies have $L>L^{\ast}$ 
as might be expected for the brightest cluster galaxy, this 
would increase the estimated redshift for each source (changing $z_{f}=3$ 
to $z_{f}=5$ will have the opposite effect). The depth of 
the optical imaging for these clusters may be inadequate to allow us to 
use VTP to detect these high-$z$ sources because fainter galaxies are 
beyond the optical limit. These objects are prime high-redshift cluster 
candidates, which we are pursuing with deep imaging in near-infrared 
bands (e.g., FLAMINGOS $J$- and $K_{s}$-band observations). At low-$z$, 
5 of the 7 sources are either near bright stars or are located on 
shallow optical images that are only complete to $i^{\prime}\sim 22$, 
which may explain their non-detection using VTP. For these we will 
seek deeper imaging. 

Our optical VTP cluster catalog, generated from the identical 6.1 square 
degree sky area covered by our {\it Chandra} fields, contains 115 sources 
with a median 
redshift of $z=0.427\pm 0.013$. Only 13 of these optical clusters are 
detected in the X-rays as an extended source, thus $89\%$ (102 out of 115) 
of our VTP detections are not included in the X-ray cluster catalog 
(see Table 4). The median redshift of the optical clusters {\it not} 
detected in the X-rays is $z=0.427\pm 0.010$. Redshift estimates are 
derived from the red-sequence filtered VTP method or measured from our 
ChaMP spectroscopic program. An example of a matched 
X-ray--optical source is presented in Figure~\ref{XOimage} for 
CXOMP~J105624.6$-$033517. This extended object also includes an X-ray 
point source, a previously known quasar at $z$=0.626 (Table 5). 
Figure~\ref{XOimage} displays the $i^{\prime}$-band optical image with 
the X-ray contours overlaid. The detection of extended emission 
in the presence of a bright point source highlights the advantage 
of {\it Chandra}'s spatial resolution in serendipitous cluster samples.

In Figure~\ref{redshift-VTPsrcs} the histogram redshift distribution of 
the optical cluster detections --- regardless of X-ray matches --- is 
depicted. The redshift distribution of these optical clusters is very 
similar to the distribution shown in Figure~\ref{redshift-his} for the 
X-ray extended sources (i.e., most of the clusters range in redshift 
from $0.2< z < 0.8$). This is expected given that the red-sequence 
filtered VTP technique provides reliable redshifts to $z\sim 0.7$ 
(see \S 3.2.1).

For the optical cluster candidates without an X-ray match, we calculated 
the X-ray upper limits (0.5-2 keV) as described previously in \S 4.2. 
Table 4 lists the coordinates, redshifts (VTP or spectroscopic), X-ray flux 
upper limits, X-ray luminosity upper limits, and {\it Chandra} PSF sizes for 
all 102 cluster candidates. 
The upper limits are $3\sigma$ values derived from background-subtracted 
counts measured within a $84\arcsec$ radius aperture (extrapolated to 
infinity using the $\beta$-model), masking out point sources and assuming 
Poisson statistics (see Figures~\ref{luminosity-his} and 
\ref{luminosity-redshift} for a comparison between the upper limits of the 
X-ray luminosity and values measured for X-ray detected extended sources).
 
The difference between the number of X-ray-to-optical matches (13 sources) 
versus optical-only detections (102 candidates) can {\it not} be explained 
as simply the difference in the spurious detection rates expected 
for both detection methods ($<10\%$ for the X-ray wavelet decomposition 
technique versus $< 20\%$ for the red-sequence filtered VTP method). A 
possible explanation is that the optical detection method is more sensitive 
to poor clusters and groups, which would be expected to contain little hot 
gas and thus be weak X-ray emitters. The dashed line in 
Figure~\ref{luminosity-his} represents the upper limit of the X-ray 
luminosity for those optical detections not matched to the extended X-ray 
source catalog. The distribution of the X-ray luminosity upper limits is 
consistent with a population of groups and ``normal'' clusters 
\citep{Mulchaey00}.  

A possible contributing factor for the disparity in the number of X-ray- 
and optical-only detections is that the X-ray counterparts to the optical 
sources have point-like X-ray emission and thus not flagged as an extended 
source. This scenario can be checked by cross-correlating the X-ray point 
source positions from ChaMP \citep{Kim04a} with the optical cluster 
candidates not matched to the extended X-ray source compilation. Of the 102 
optical clusters without extended X-ray matches, 9 are detected by ChaMP 
as X-ray point sources. We derive extended source flux upper limits for 
these, as described in \S 4.2, but we exclude the X-ray point source 
(conservatively using the 90\% encircled counts radius). The resulting 
median X-ray luminosity of these 9 upper limits is 
$(3.87\pm 12.74)\times 10^{42}~\mbox{ergs}~\mbox{s}^{-1}$. Thus, 
it is reasonable to assume that $\sim 9\%$ (9 out of 102) of the 
optical-only cluster candidates are not detected as extended X-ray sources 
because their X-ray emission is point-like, thus excluding them from our 
extended X-ray source catalog. To check whether the {\it Chandra} PSF size 
has a significant impact on the fraction of X-ray-to-optical detections, we 
plot in Figure~\ref{PSFsize-histogram} the histogram of sources by 
{\it Chandra} PSF size both for the sample of optical clusters not detected 
as an extended X-ray source and for all 55 extended X-ray sources. 
Figure~\ref{PSFsize-histogram} gives 
the visual appearance that the two distributions are consistent. The 
two-sample Kolmogorov-Smirnov test yields D=0.14, with a 55\% 
probability that the null hypothesis can not be rejected 
(i.e., confirming that the histograms are not inconsistent). 
A small number of truly extended sources may also be missed if they 
host bright embedded point sources. Both detailed simulations and a 
larger {\it Chandra}-selected cluster sample would help to constrain 
this contribution. The ChaMP has initiated such a study, which is beyond 
the scope of the current paper.

Finally, there is the possibility that some fraction of the optical-only 
detections are due to the chance alignment of filaments in the large-scale 
structure of the cosmic web of galaxies \citep{Gladd00}. Since our optical 
cluster candidates  are detected by filtering with respect to the 
color-magnitude red-sequence, we minimize false optical detections. In 
Table 6, we tabulate information available from NED regarding sources 
located near each optical source not detected in the X-rays.  

\subsubsection{Optical Cluster Richness}

Galaxy cluster richness is an important characteristic that provides 
information on cluster mass \citep[e.g.,][]{McNamara01}. Historically, 
cluster richness has been described using the Abell Richness Class (ARC), 
first defined by \cite{Abell58}. Several studies have shown that the 
Abell richness parameter is not a well-defined quantity and is subjected 
to numerous observational biases, including projection effects 
\citep[e.g.,][]{Lucey83,vanHaarlem97,Miller99}. 
For our cluster sample, we have elected to use the parameter $B_{gc}$ 
\citep{Yee99,Yee03} to characterize optical cluster richness. The 
$B_{gc}$ parameter is a measure of the cluster center--galaxy correlation 
amplitude and is related to the correlation function defined by; 
$\xi (r)=B_{gc}r^{-\gamma}$ \citep{Longair79}. Observationally, it is 
easier to compute the angular correlation function, which can be 
approximated using a power-law of the form 
$w(\theta)=A_{gg}\theta^{1-\gamma}$ \citep[see][]{Davis83} rather than 
$\xi (r)$. Determining the angular distribution of galaxies about the 
cluster center provides a measure of $A_{gg}$, the number of 
background-subtracted galaxies within some angular radius $\theta$ of the 
adopted cluster center. The amplitude of the angular correlation function 
can be expressed as (assuming a fixed $\gamma$) 
$A_{gc}=(N_{net}/N_{bg})[(3-\gamma)/2]\theta^{\gamma-1}$, where $N_{net}$ 
is the background-corrected galaxy counts, and $N_{bg}$ is the background 
counts \citep{Yee99}. Assuming spherical symmetry, $A_{gc}$ and $B_{gc}$ 
can be related \citep{Longair79} via:

\begin{equation}
B_{gc}=N_{bg}\frac{D^{\gamma -3}A_{gc}}{I_{\gamma}\Psi [M(m_{o},z)]},
\end{equation}

\noindent where $N_{bg}$ is the background galaxy counts measured to 
apparent magnitude $m_{o}$, $D$ is the angular-diameter distance, 
$I_{\gamma}$ is an integration constant, and $\Psi [M(m_{o},z)]$ is the 
integrated luminosity function to absolute magnitude $M(m_{o},z)$ 
corresponding to the apparent magnitude limit $m_{o}$ at the cluster 
redshift. 

The uncertainty of $B_{gc}$ can be computed from \citep{Yee99}:

\begin{equation}
\frac{\Delta B_{gc}}{B_{gc}}=\frac{(N_{net}+1.3^{2}N_{bg})^{1/2}}{N_{net}},
\end{equation}

\noindent where the factor of $1.3^{2}$ accounts for the field-to-field 
fluctuation of background galaxy counts above the expected Poisson 
distribution \citep[e.g.,][]{Yee87,Lopez97,Barkhouse03}. 

The $B_{gc}$ parameter has been used in numerous studies to quantify 
the environment of quasars and radio galaxies 
\citep{Yee84,Yee87,Prestage88,Yates89,Ellingson91}, BL Lacertae objects 
\citep{Smith95,Wurtz97}, Seyfert galaxies \citep{DeRobertis98}, Abell 
clusters \citep{Anderson94,Lopez97,Yee99,Barkhouse03}, {\it ROSAT} 
clusters \citep{Gilbank04}, and  RCS clusters \citep{Gladd00,Hicks04}.  

The primary steps associated with measuring $B_{gc}$ for our cluster sample 
involve counting galaxies to a fixed absolute magnitude within a fixed 
physical radius of the adopted cluster center. The galaxy counts are then 
background-corrected using the luminosity function generated from 
four randomly-positioned deep ChaMP optical images that do not contain 
known clusters or bright stars. Galaxies are counted within a radius 
of $0.5h^{-1}_{70}~\mbox{Mpc}$ from the cluster center. The cluster 
center is defined as the centroid of the brightest cluster galaxy or the 
center of the galaxy density enhancement if no obvious BCG exists. 
Following the procedure of \citet{Yee99}, we include only those galaxies 
brighter than the $K$- and evolution-corrected value 
($Q=-1.4z$; Yee \& L\'{o}pez-Cruz 1999) of $M^{\ast}+2$, where 
$M^{\ast}$ is the value of $M^{\ast}_{i^{\prime}}=-21.52$ as 
implemented for our 
red-sequence filtered VTP technique (see \S 3.2.1). In addition to the 
general procedure outlined in \citet{Yee99}, we only include galaxies that 
have $r^{\prime}-i^{\prime}$ colors within $\pm 0.5$ mag of the cluster 
red-sequence for our specific target (identical color cuts were also 
applied to the background galaxy population). This color-selection step 
helps to minimize uncertainties in the luminosity function and galaxy 
evolutionary corrections at $z\sim 0.4$ (the median redshift of our sample) 
by statistically selecting red-sequence early-type galaxies with known 
properties (see \citealt{Hicks04} for a similar application).

In Figure~\ref{Bgc-lum} we plot $L_X$ versus $B_{gc}$ for three cluster 
samples; 1) the cluster sample with X-ray and optical matches (solid 
circles), 2) the optical-only clusters detected by VTP (arrows), and 
3) a sample of 35 Abell clusters (solid triangles) from \citet{Barkhouse03}. 
We include the Abell cluster sample to provide a reference of the expected 
range of richness of known clusters. The X-ray luminosity values for the 
Abell cluster sample are taken from \citet{Ledlow03} for the 0.5-2 keV 
energy band and transformed to our cosmology. It is apparent from the 
figure that $B_{gc}$ is weakly correlated with $L_X$ such that richer 
clusters (as denoted by larger values of $B_{gc}$) are more luminous in 
the X-rays. The large scatter between these measurements has also been 
seen in previous studies \citep{Yee03,Gilbank04,Hicks04}. Due to the 
relatively bright optical magnitude limits compared to 
$m_{i^{\prime}}^{\ast}$ for three fields, 1 cluster from the 
X-ray-optical matched sample and 8 from the optical-only group are not 
included in the $B_{gc}$ analysis since galaxies could not be counted to 
the adopted absolute magnitude limit.

The $B_{gc}$ values for all three groups of clusters have been measured in 
a consistent manner using the same cosmology and selection of galaxies 
relative to the individual cluster red-sequence. The median value 
of $B_{gc}$ for the three samples are; $B_{gc}^{med}=1849\pm 236$ 
for the group of clusters with X-ray and 
optical counterparts, $B_{gc}^{med}=944\pm 47$ for the optical-only 
detected clusters, and $B_{gc}^{med}=1444\pm 114$ for the 35 Abell clusters.  

The median richness of the optical-only clusters is $3.8\sigma$ lower 
than the value measured for the matched X-ray/optical sample. This 
provides additional support to the conclusions drawn from 
Figure~\ref{luminosity-his} and~\ref{luminosity-redshift} that 
the optical-only cluster sample consists mainly of poor clusters and 
groups that are too X-ray weak to be detected in our sample. However, 
there is evidence in Figure~\ref{Bgc-lum} that some of the optical 
sources not detected in the X-ray have optical richness consistent with 
Abell-like clusters (see \S 5.0).

\section{Discussion and Conclusions}

The primary goal of this paper is to present the X-ray and optical 
properties of a compilation of extended X-ray sources discovered 
serendipitously as part of the ChaMP. The availability of $\sim 13$ 
square degrees of {\it Chandra} archival data with deep mosaic 
optical coverage from 56 NOAO/4m fields allows us to test independently 
cluster detection schemes in the X-ray and optical passbands. By 
cross-correlating the resulting compilations from the overlapping 
6.1 square degrees, we are able to extend the X-ray and optical 
analysis of clusters/groups to lower X-ray 
luminosities than previous {\it ROSAT} cluster surveys such as 
REFLEX (see Figure~\ref{luminosity-redshift}). The 160 Square Degree 
{\it ROSAT} Survey provides a better match to our sample than the 
REFLEX survey. The 160 Square Degree Survey covers the same range in 
redshift and $L_X$ as the ChaMP cluster survey but includes a larger 
fraction of sources at lower redshift.

From our sample of 55 extended X-ray sources, we measured a matched 
fraction of $46\%$ (13 out of 28) with cluster candidates detected by our 
red-sequence filtered VTP method. This matched fraction only includes 
those sources that are located in areas that have overlapping X-ray 
and optical coverage, and excludes the six X-ray sources that 
are coincident with single bright galaxies in the optical. The $46\%$ 
matched fraction is not too different from typical values measured 
in other X-ray-optical studies. \citet{Donahue02} found that 
$60\%$ (26 out of 43) of their {\it ROSAT}-detected clusters have 
optical counterparts as measured by their matched-filter algorithm. This 
is similar to \citet{Kolokotronis06} who matched $68\%$ (13 out of 19) 
of their {\it XMM-Newton} sources with optical clusters detected using 
a smoothing percolation method. We note, however, that a direct 
comparison to other studies is problematic given that the matched 
fraction is expected to depend on the flux limit achieved in each passband 
and the X-ray-to-optical flux ratio for specific types of galaxy 
clusters.

Since there is no such thing as the ``perfect'' cluster 
detector, each cluster-finding method is subject to bias. The X-ray 
wavelet decomposition technique relies on the ability to separate 
extended- and point-sources. This task becomes more difficult the 
larger the off-axis angle and can prevent the inclusion of extended 
X-ray sources that are found near the field edge. In addition, poor 
clusters or groups with shallow potential wells may only emit detectable 
X-rays from regions smaller than the {\it Chandra} PSF. The 
red-sequence filtered VTP optical cluster detection technique relies on the 
presence of the early-type galaxy red-sequence to help 
improve the contrast of clusters above the field galaxy population. For 
increasing redshift, the fraction of blue cluster galaxies (usually 
associated with later types) has been found to increase \citep[the well-known 
Butcher-Oemler effect;][]{Butcher84}. The increase in the blue 
fraction will have an effect on the efficiency of any optical 
cluster finder that relies on the existence of the cluster 
red-sequence \citep[e.g.,][]{Gladd00,Donahue02}. 

\begin{figure}
\figurenum{9}
\plotone{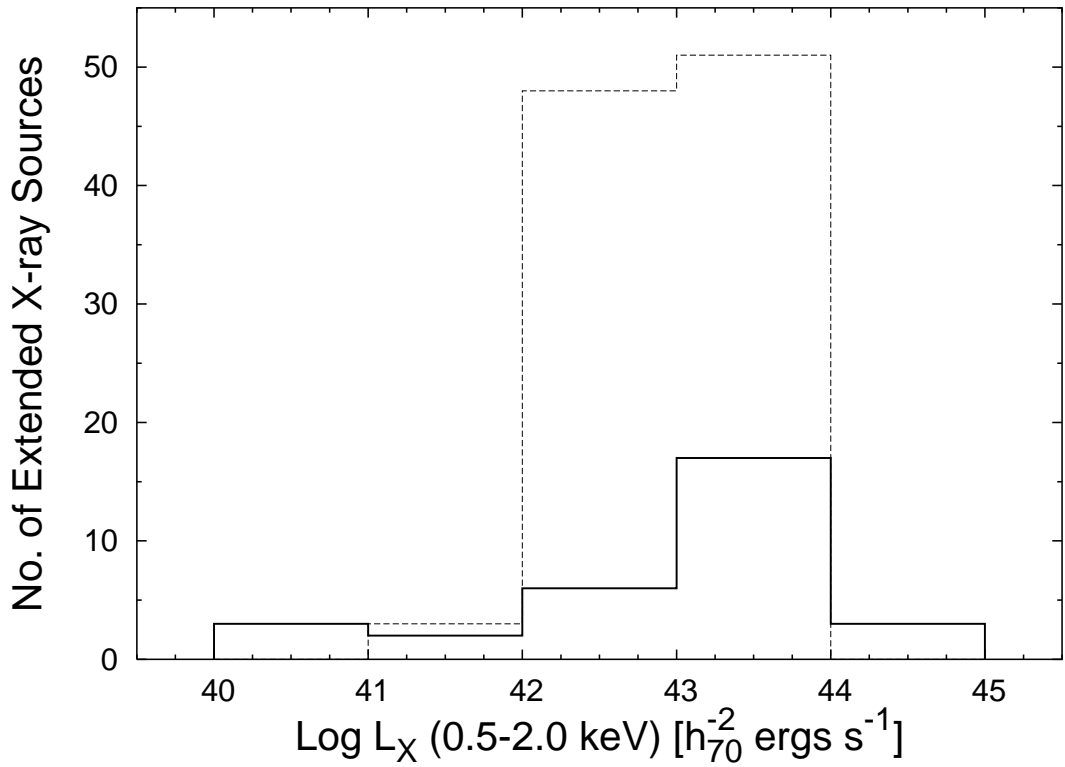}
\caption{The solid line depicts the distribution of X-ray luminosity for 
31 serendipitously-detected extended X-ray sources. The dashed line 
shows the distribution of upper limits to the X-ray luminosity for 
102 cluster candidates detected by VTP but not in the X-rays.}
\label{luminosity-his}
\end{figure}

\begin{figure}
\figurenum{10}
\plotone{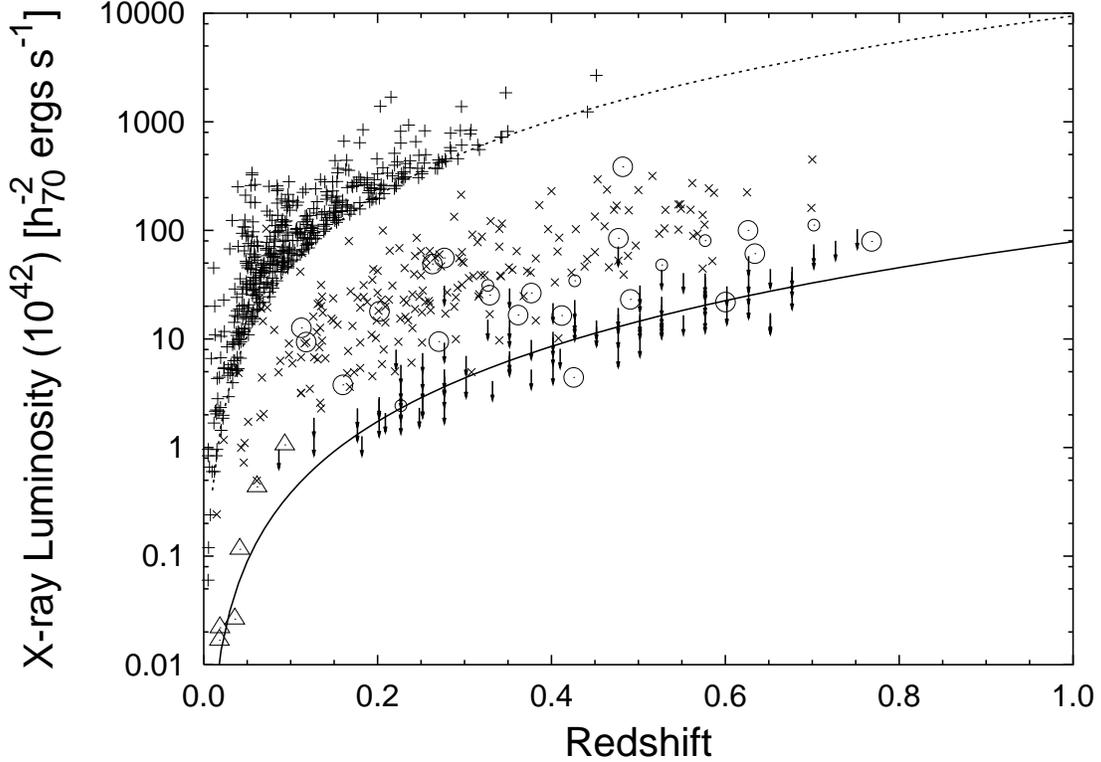}
\caption{The distribution of X-ray luminosity (0.5-2 keV) as a function of 
redshift. The open circles mark the 25 extended X-ray sources associated 
with clusters with redshifts, while the open triangles depict the 6 low-$z$ 
galaxies. The arrows represent X-ray luminosity upper limits for the 102 
sources detected by VTP but undetected in the X-rays. The ``+'' symbols depict 
the 447 X-rays clusters from the REFLEX compilation, while 200 clusters from 
the 160 Square Degree {\it ROSAT} Survey is represented by ``x'' symbols. The 
solid line indicates the flux limit of 
$1.5\times 10^{-14}~\mbox{erg}~\mbox{s}^{-1}~\mbox{cm}^{-2}$ (0.5-2 keV) 
which is shown for illustrative purposes. The flux limit of the REFLEX 
sample (converted to the 0.5-2 keV energy band) is shown as the dashed line, 
corresponding to 
$1.8\times 10^{-12}~\mbox{erg}~\mbox{s}^{-1}~\mbox{cm}^{-2}$. Larger 
symbols represent sources with spectroscopic redshifts while sources with 
VTP-only estimated redshifts are marked with smaller symbols.}
\label{luminosity-redshift}
\end{figure}

\begin{figure}
\figurenum{11}
\epsscale{1.0}
\plotone{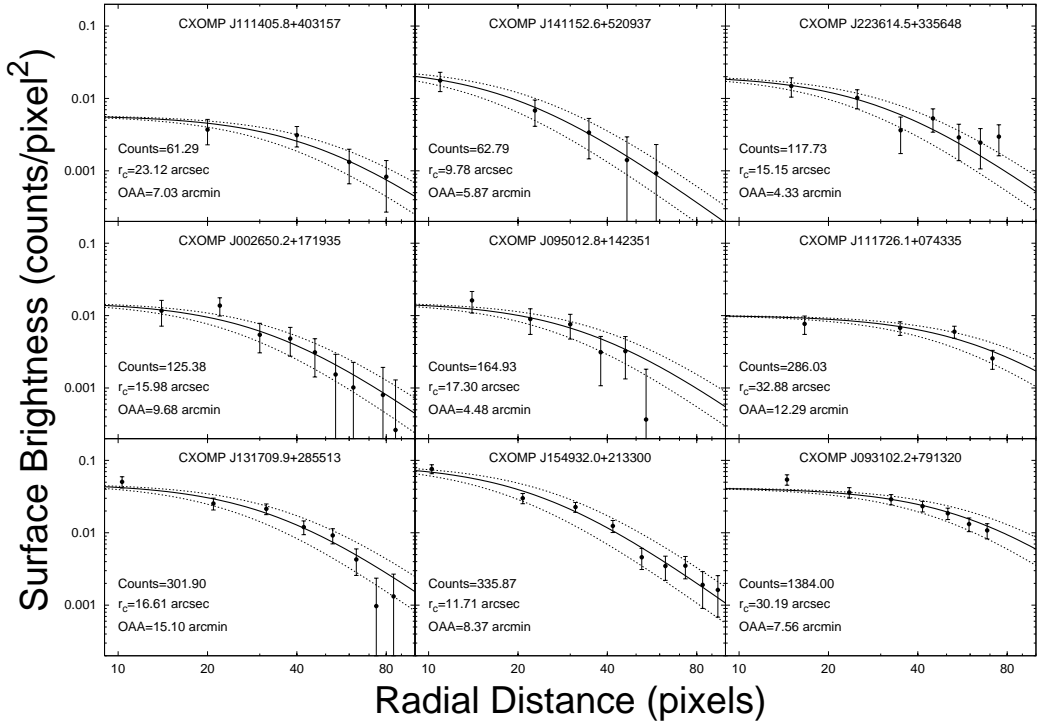}
\caption{The X-ray surface brightness radial profiles for a representative 
sample of nine extended sources. The best-fit $\beta$-model ($\beta=0.67$; 
solid line) along with the associated $1\sigma$ uncertainty (dashed lines) 
are presented for each source. The depicted sources range in total counts 
from 61.29 to 1384 and core radii extending from 9.78 to 32.88 arcsec. 
The total counts, core radius, and off-axis angles are tabulated in Table 4 
for each extended X-ray source.}
\label{SB-radial}
\end{figure}

\begin{figure}
\figurenum{12}
\plotone{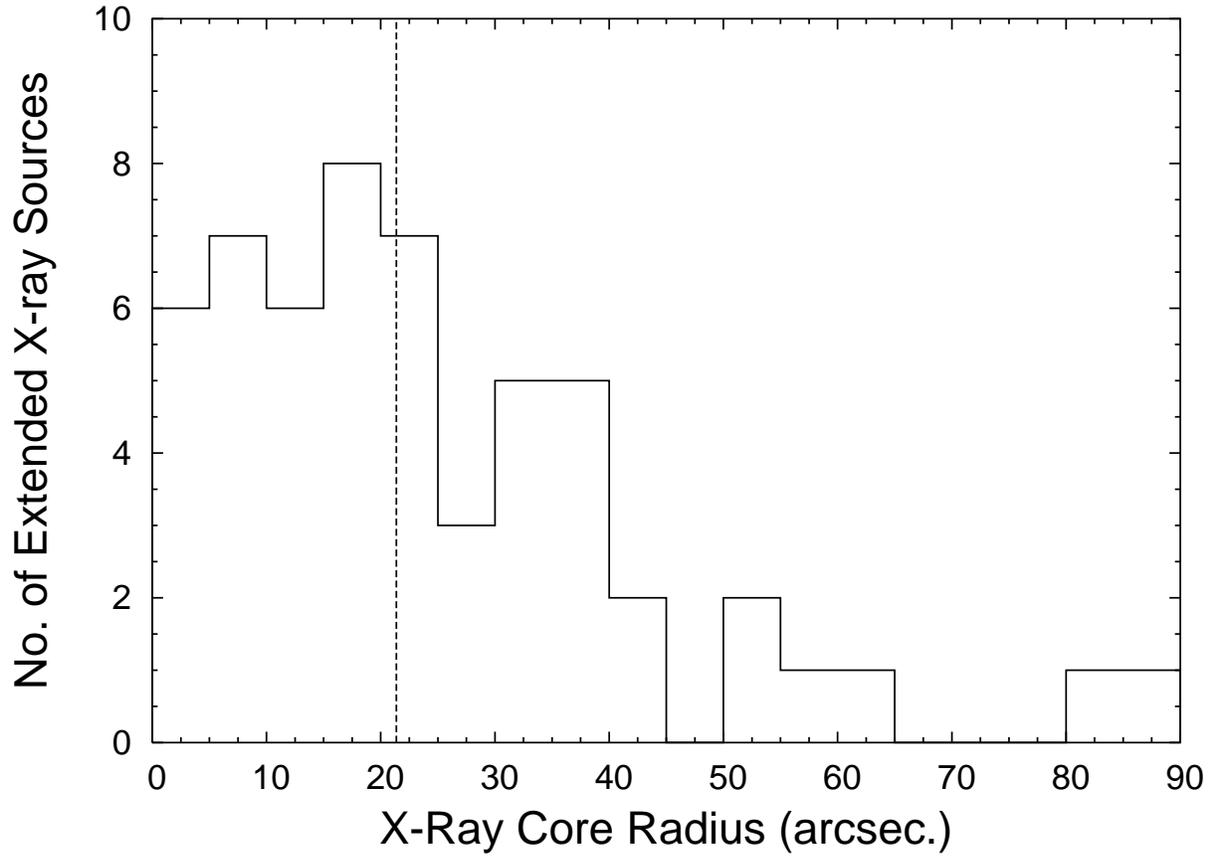}
\caption{The histogram distribution of the measured core radius for 55 
extended X-ray sources. The median core radius of $21.37\arcsec$ is 
represented by the vertical dashed line.}
\label{coreradius-his}
\end{figure}

\begin{figure}
\figurenum{13}
\epsscale{0.8}
\plotone{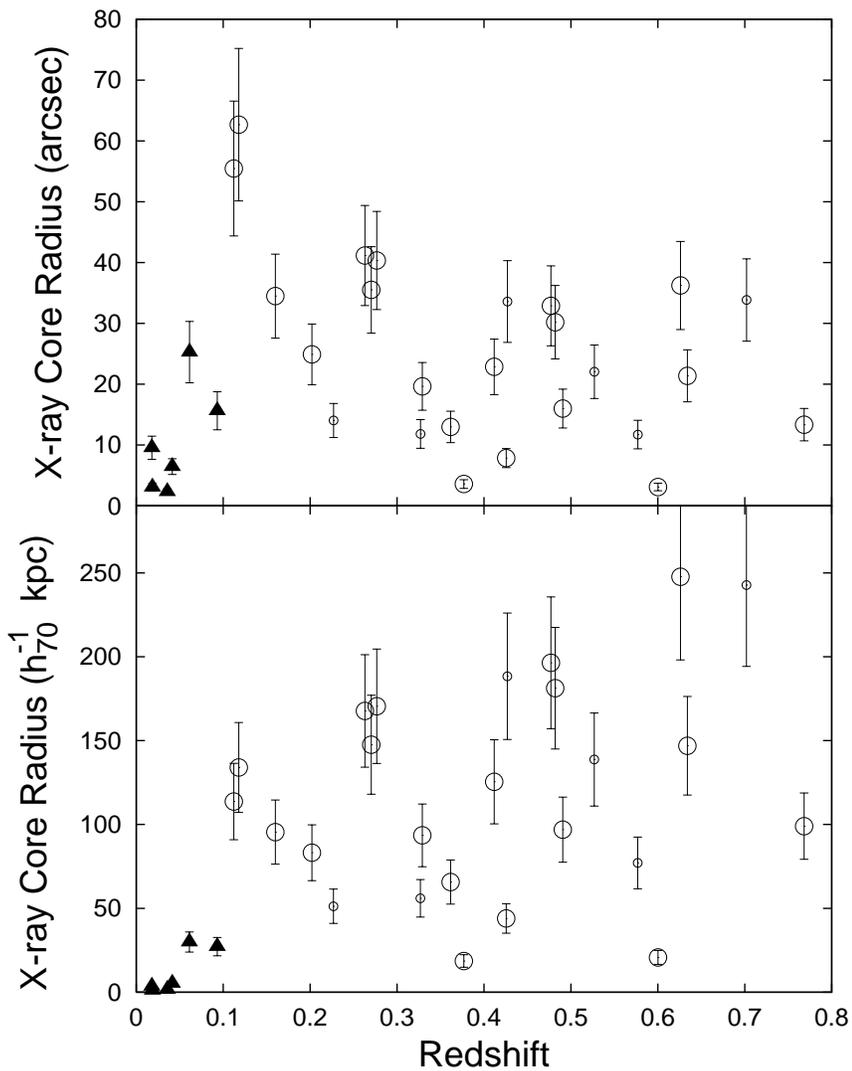}
\caption{The angular (arcsec) and physical core radius ($h^{-1}_{70}$ kpc) 
for 31 extended X-ray sources as a function of redshift. The solid triangles 
depict extended X-ray sources associated with single galaxies and open 
circles represent X-ray-detected clusters; large symbol size indicate 
sources with spectroscopic redshifts while the smaller symbols represent 
objects with red-sequence filtered VTP redshifts. No obvious correlation is 
apparent between the core radius and redshift.}
\label{coreradius-redshift}
\end{figure}

\begin{figure}
\figurenum{14}
\epsscale{1.0}
\plotone{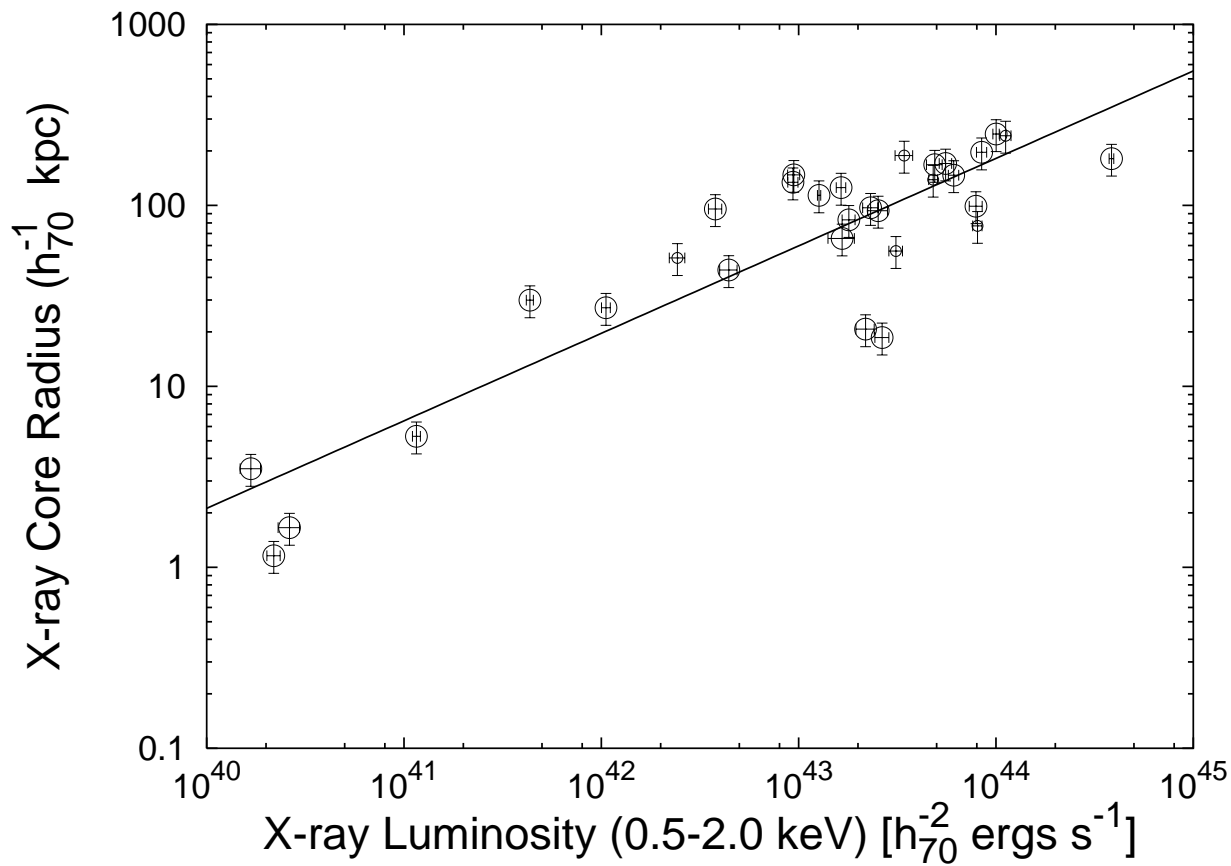}
\caption{The metric core radius ($h^{-1}_{70}$ kpc) as a function of 
X-ray luminosity for the 31 extended X-ray sources with estimated 
redshifts. The core radius and luminosity are well correlated with a 
best-fit power-law (solid line) of 
$r_c\propto L_X^{0.48\pm 0.04}$. Sources with 
spectroscopic redshifts are depicted with the large symbols and objects 
with VTP estimated redshifts are shown with the smaller symbols.} 
\label{coreradius-flux}
\end{figure}

\begin{figure}
\figurenum{15}
\epsscale{1.0}
\plotone{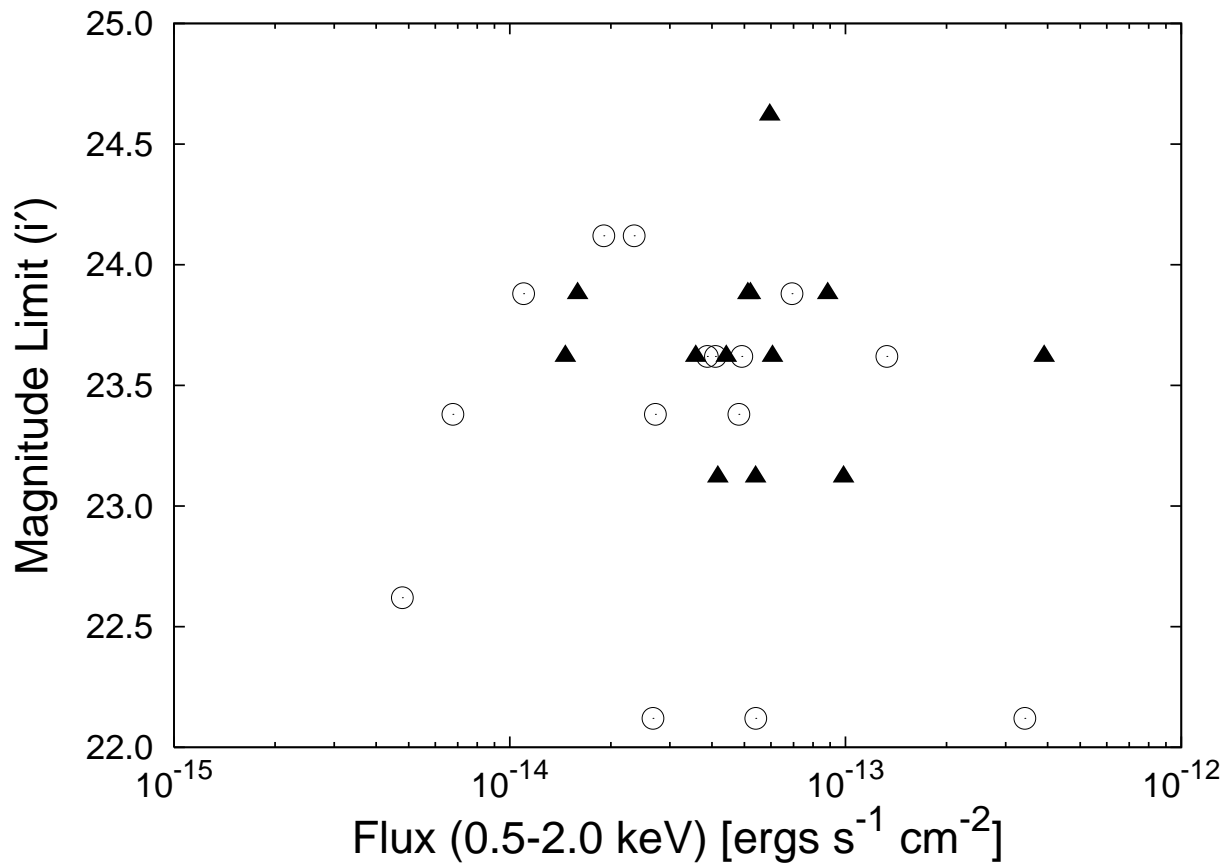}
\caption{The $i^{\prime}$-band optical field turnover magnitude versus 
X-ray flux for the 28 extended X-ray sources that overlap with the mosaic 
imaging (the 6 low-$z$ galaxies are excluded). The 13 matched optical/X-ray 
sources are depicted as the solid triangles while the 15 X-ray sources without 
optical VTP matches are represented by the open circles. Several non-VTP 
detected sources with faint optical magnitude field limits and bright 
$f_{X}$ are prime candidates for deeper optical/near-IR follow-up imaging 
to detect high redshift clusters.} 
\label{iMag-Flux}
\end{figure}

\begin{figure}
\figurenum{16}
\plotone{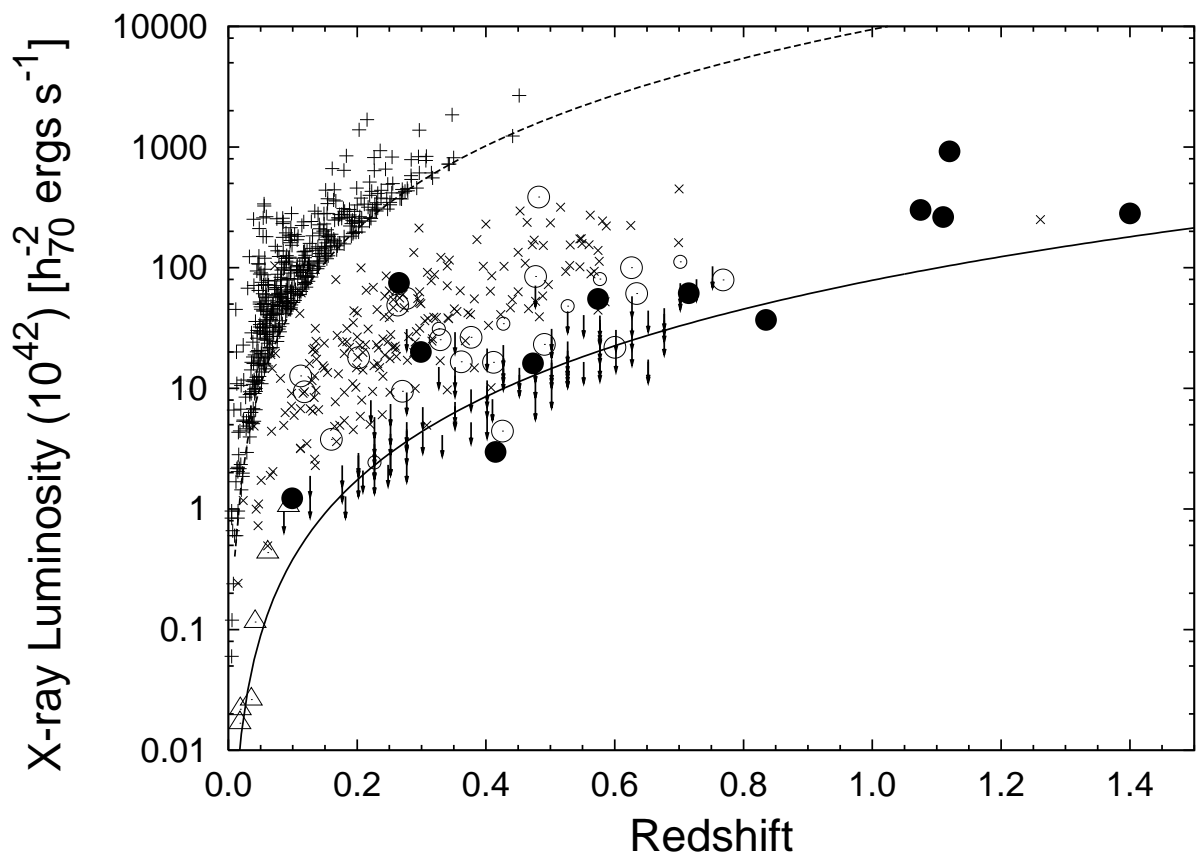}
\caption{The distribution of X-ray luminosity (0.5-2 keV) as a function of 
redshift. This plot is similar to Figure~\ref{luminosity-redshift} and 
includes 12 extended X-ray sources that have no VTP-detected optical 
counterparts or measured redshifts (solid circles). Redshifts for these 
sources have been estimated using the $i^{\prime}$-band magnitude of the 
galaxy nearest to the X-ray centroid.}
\label{luminosity-redshift3}
\end{figure}

\begin{figure}
\figurenum{17}
\epsscale{0.9}
\plotone{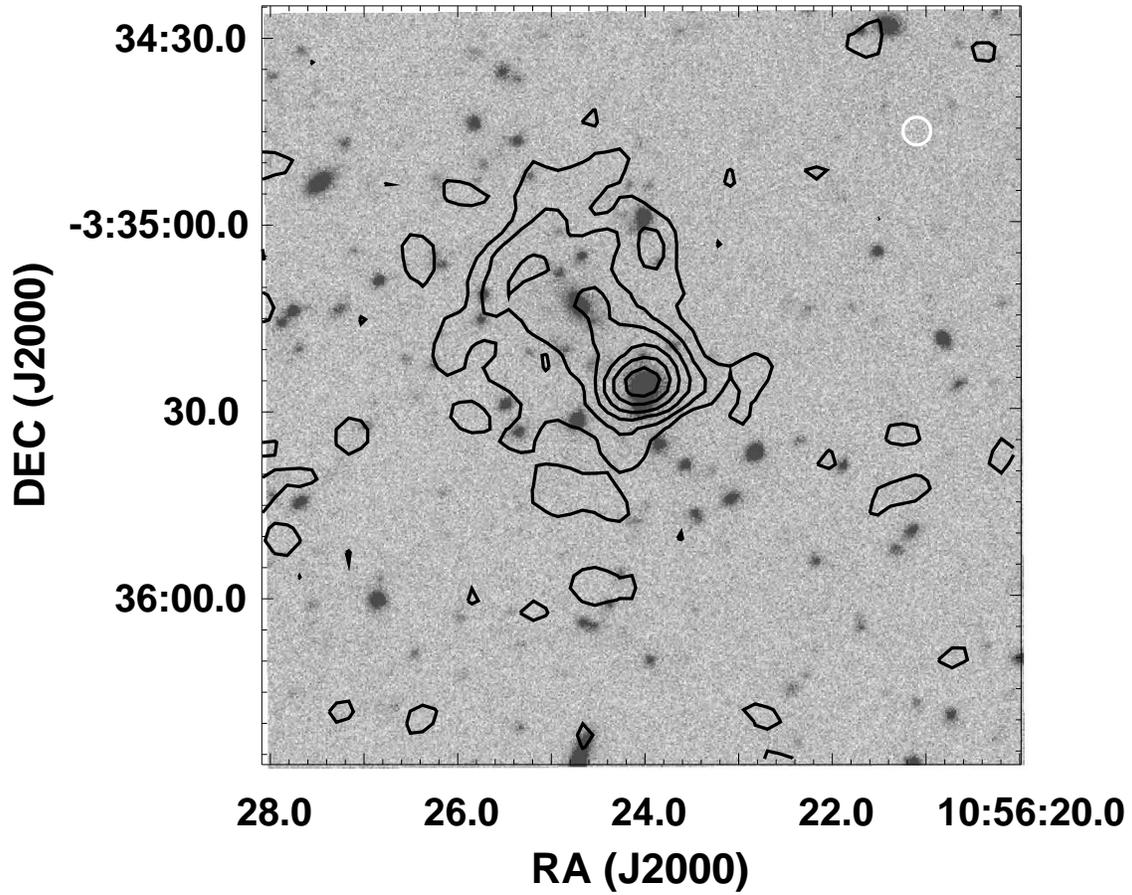}
\caption{Optical $i^{\prime}$-band image of extended X-ray source 
CXOMP~J105624.6$-$033517 (see Table 1) with the X-ray contours overlaid. 
This is an example of an extended X-ray source with an X-ray point-source 
(a known $z$=0.626 quasar) embedded within. The size of the {\it Chandra} 
PSF at the source off-axis angle ($8.8\arcmin$) is $\sim 4.5\arcsec$ and 
is represented by the white circle in the upper-right corner of the figure.}
\label{XOimage}
\end{figure}

\begin{figure}
\figurenum{18}
\plotone{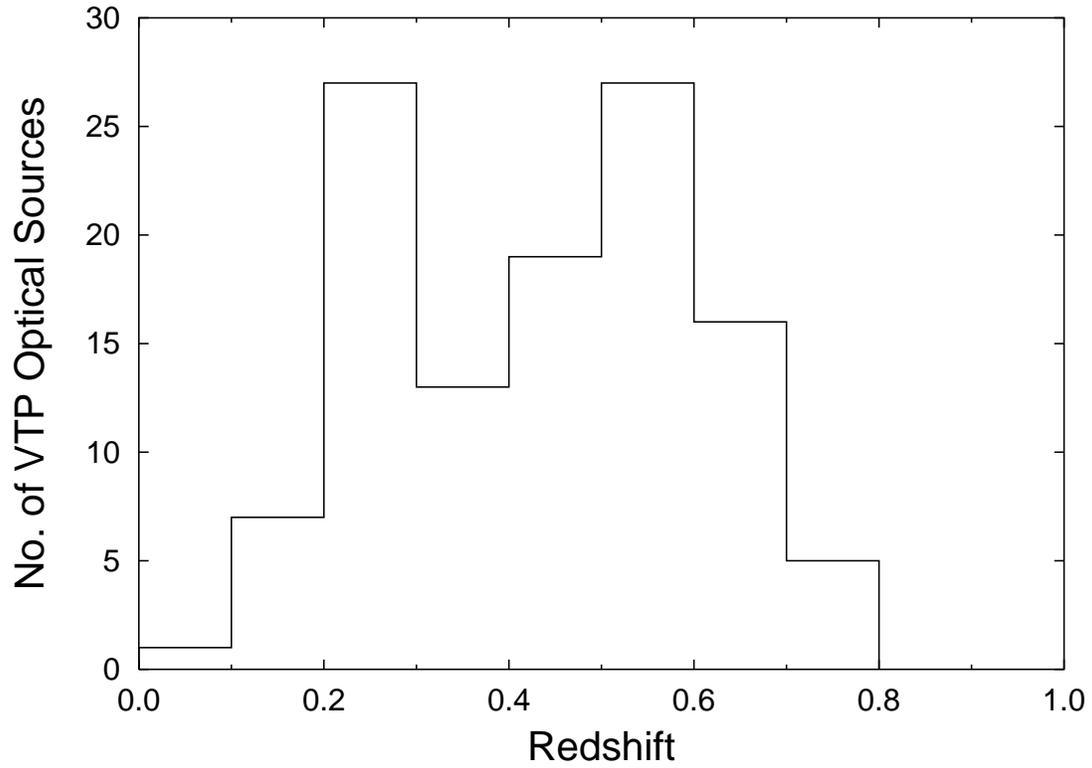}
\caption{The redshift distribution of 115 red-sequence filtered VTP optical 
detections of which 13 have X-ray counterparts. The VTP redshift estimates 
are based on the redshift of the associated red-sequence slice that 
maximizes the confidence and probability of being a real cluster. When 
available, spectroscopic redshifts are used in lieu of the VTP estimates.}
\label{redshift-VTPsrcs}
\end{figure}

\begin{figure}
\figurenum{19}
\plotone{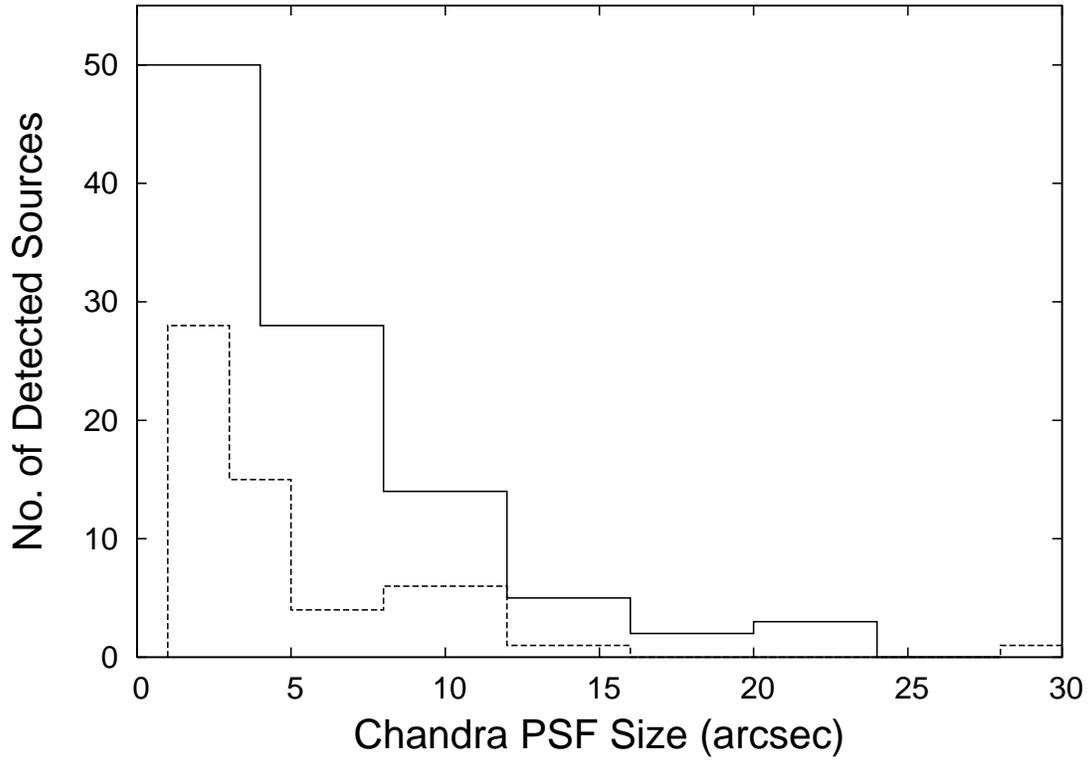}
\caption{A comparison of the histogram distribution of 
detected candidate cluster sources as a function of the {\it Chandra} 
PSF size at the source position. The solid line depicts the distribution 
of the 102 optical VTP detections that were not detected as extended 
X-ray sources, while the dashed line represents the 55 extended X-ray 
sources. A K-S test demonstrates that the two distributions are not 
inconsistent with each other.} 
\label{PSFsize-histogram}
\end{figure}

\begin{figure}
\figurenum{20}
\plotone{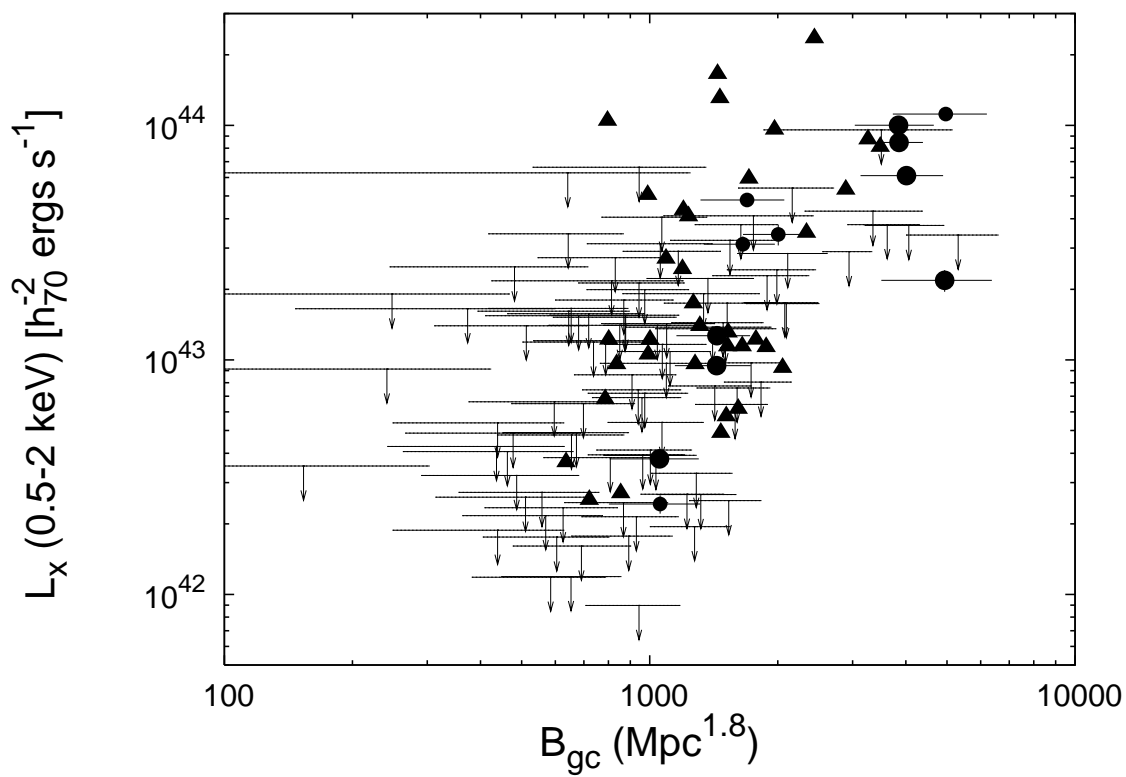}
\caption{The X-ray luminosity (0.5-2 keV) versus the richness parameter 
$B_{gc}$ for three cluster samples. The matched X-ray-optical cluster 
sample is depicted with the solid circles while the optical-only clusters are 
represented by the arrows (upper limits for $L_X$). Clusters 
with spectroscopic redshifts are plotted with larger symbol size while 
sources with VTP redshifts are shown with the smaller symbols. For 
comparison purposes, a sample of 35 Abell clusters is shown by the solid 
triangles. All uncertainties are $1\sigma$ values.}
\label{Bgc-lum}
\end{figure}

A comparison of the sample of optically-selected VTP clusters with those 
detected in the X-rays yields a matched fraction of $11\%$ (i.e., only 
13 out of 115 optical clusters have a detected X-ray counterpart). As 
discussed in \S 4.3, many of these sources have X-ray upper limits that 
are consistent for a population of groups and ``normal'' clusters (see 
Figure~\ref{luminosity-his}). A comparison of the richness of the 
optical-only versus X-ray-optically matched cluster samples (see 
Figure~\ref{Bgc-lum}) shows that the average richness of the optical-only 
VTP sample is smaller than the cluster sample with X-ray and optical 
counterparts by $\sim 4\sigma$. This result supports the hypothesis 
that the optical-only sample 
is either (1) composed mainly of poor systems that are undetected by our 
X-ray observations due to the lack of sufficient hot intracluster gas 
or, (2) contaminated by non-virialized filaments of the large-scale 
structure. In addition, the comparison of optical richness as characterized 
by $B_{gc}$ (see Figure~\ref{Bgc-lum}) shows the presence of several 
clusters not detected in our X-ray data that have an optical richness similar 
to Abell-type clusters. We examined the possible impact that the X-ray 
exposure time has on the detection of these systems by looking at the median 
exposure times for the VTP-only detected sources with $B_{gc}>3000$ 
(5 sources) and those X-ray detected sources with  $B_{gc}>3000$  (also 5 
sources). The median vignetting-corrected X-ray exposure time of the 
VTP-only sources is $19.5\pm 12.7~\mbox{ksec}$, where the uncertainty 
is the dispersion. For the X-ray detected sources we find a median value 
of $44.9\pm 12.4~\mbox{ksec}$. The median X-ray exposure times of the 
positions of the VTP-only detected clusters is at most $1.4\sigma$ 
lower than the X-ray detected extended sources. Including clusters of 
lower $B_{gc}$ further decreases the significance of this difference 
($0.5\sigma$ for $B_{gc}>2000$). This analysis indicates that it is 
unlikely that many of the VTP-only detected optical sources are missed 
in the X-rays due to shallow X-ray exposure times.

To increase the cluster sample size and search for rare luminous X-ray 
clusters, we will extend this study to include archival images from 
{\it Chandra} Cycles $3-6$. We will use primarily SDSS photometry for 
optical coverage, and a red-sequence VTP method to discriminate cluster 
red-sequences out to $z\sim 1.1$, or $z\sim 0.5$ if we only include galaxies 
brighter than $m^{\ast}_{z^{\prime}}$ (based on the optical magnitude 
limit of the SDSS $z^{\prime}$-band data; i.e. $z^{\prime}_{lim}\sim 20.5$).

\acknowledgments

We gratefully acknowledge support for this project through NASA under 
CXC archival research grant AR4-5017X. RAC, PJG, DWK, AEM, HT, and BJW 
also acknowledge support through NASA contract NAS8-39073 (CXC). The 
authors thank the staff at CTIO, KPNO, and Las Campanas for assistance 
with the optical observations presented here.

We thank Taddy Kodama for use of his cluster red-sequence models, and 
Howard Yee for discussions and access to his $B_{gc}$ code. We also thank 
the referee for helpful comments and suggestions.

\clearpage

\begin{deluxetable}{ccrrcc}
\tablecaption{X-RAY-DETECTED EXTENDED SOURCES}
\tabletypesize{\small}
\tablewidth{0pt}
\tablehead{
\colhead{Source Name} &\colhead{ObsID} &\colhead{RA} &\colhead{DEC}&\colhead{Exp\tablenotemark{a}} 
&\colhead{Galactic $\mbox{N}_{\mbox{H}}$\tablenotemark{b}}\\ 
&&\colhead{(J2000)} &\colhead{(J2000)} &\colhead{(sec)} 
&\colhead{($10^{20}~\mbox{cm}^{-2}$)}}
\startdata
CXOMP~J002650.2+171935 & 929 & 00 26 50.2 & +17 19 35.7 & 40346 & 4.19 \\
CXOMP~J005848.1$-$280035 & 2248 & 00 58 48.1 & $-$28 00 35.7 & 12138 & 1.55 \\
CXOMP~J010214.1+314915 & 521 & 01 02 14.1 & +31 49 15.6 & 54166 & 5.50 \\
CXOMP~J010607.0+004943 & 2180 & 01 06 07.0 & +00 49 43.7 & 3757 & 3.15 \\
CXOMP~J010610.3+005126 & 2180 & 01 06 10.3 & +00 51 26.1 & 3757 & 3.15 \\
CXOMP~J013642.6+204843 & 2129 & 01 36 42.6 & +20 48 43.7 & 45094 & 5.71 \\
CXOMP~J033639.4$-$045515 & 796 & 03 36 39.4 & $-$04 55 15.4 & 60512 & 4.98 \\
CXOMP~J033722.6$-$045906 & 796 & 03 37 22.7 & $-$04 59 05.8 & 60512 & 4.98 \\
CXOMP~J033755.1$-$050733 & 796 & 03 37 55.1 & $-$05 07 33.6 & 60512 & 4.98 \\
CXOMP~J033757.8$-$050001 & 796 & 03 37 57.8 & $-$05 00 00.9 & 60512 & 4.98 \\
CXOMP~J040351.2$-$170823 & 2182 & 04 03 51.2 & $-$17 08 23.2 & 3891 & 2.30 \\
CXOMP~J054152.7$-$410702 & 914 & 05 41 52.7 & $-$41 07 02.7 & 51050 & 3.59 \\
CXOMP~J054240.1$-$405503 & 914 & 05 42 40.1 & $-$40 55 03.3 & 51050 & 3.59 \\
CXOMP~J063057.7+820701 & 1602 & 06 30 57.7 & +82 07 01.2 & 47933 & 5.27 \\
CXOMP~J090634.4+340055 & 1596 & 09 06 34.4 & +34 00 55.6 & 9907 & 2.28 \\
CXOMP~J091008.4+541852 & 2227 & 09 10 08.4 & +54 18 52.3 & 107136 & 1.98 \\
CXOMP~J091126.6+055012 & 419 & 09 11 26.6 & +05 50 12.5 & 29165 & 3.70 \\
CXOMP~J091301.4+054814 & 419 & 09 13 01.4 & +05 48 14.0 & 29162 & 3.70 \\
CXOMP~J093102.2+791320 & 839 & 09 31 02.2 & +79 13 20.9 & 19165 & 1.90 \\
CXOMP~J093352.9+552619 & 805 & 09 33 52.9 & +55 26 19.6 & 41296 & 1.99 \\
CXOMP~J095012.8+142351 & 2095 & 09 50 12.8 & +14 23 51.7 & 13962 & 3.13 \\
CXOMP~J101008.7$-$124013 & 926 & 10 10 08.7 & $-$12 40 13.1 & 44730 & 6.74 \\
CXOMP~J101115.3$-$124147 & 926 & 10 11 15.3 & $-$12 41 47.1 & 44733 & 6.74 \\
CXOMP~J105624.6$-$033517 & 512 & 10 56 24.6 & $-$03 35 17.4 & 90211 & 3.67 \\
CXOMP~J111405.8+403157 & 2209 & 11 14 05.8 & +40 31 57.4 & 30054 & 1.91 \\
CXOMP~J111726.1+074335 & 363 & 11 17 26.1 & +07 43 35.3 & 26832 & 4.01 \\
CXOMP~J111730.2+074618 & 363 & 11 17 30.2 & +07 46 18.7 & 26832 & 4.01 \\
CXOMP~J114008.2$-$263132 & 898 & 11 40 08.2 & $-$26 31 32.6 & 39978 & 4.96 \\
CXOMP~J114118.8+660209 & 536 & 11 41 18.8 & +66 02 09.4 & 119222 & 1.18 \\
CXOMP~J122927.1+752037 & 2253 & 12 29 27.1 & +75 20 37.2 & 48010 & 2.69 \\
CXOMP~J122940.6+752106 & 2253 & 12 29 40.6 & +75 21 06.6 & 48010 & 2.73 \\
CXOMP~J131709.9+285513 & 2228 & 13 17 09.9 & +28 55 13.7 & 112806 & 1.04 \\
CXOMP~J131722.0+285353 & 2228 & 13 17 22.0 & +28 53 53.0 & 112806 & 1.04 \\
CXOMP~J134507.8+000359 & 2251 & 13 45 07.8 & +00 03 59.0 & 9760 & 1.93 \\
CXOMP~J134514.6$-$000846 & 2251 & 13 45 14.6 & $-$00 08 46.5 & 9760 & 1.93 \\
CXOMP~J141152.6+520937 & 2254 & 14 11 52.6 & +52 09 37.2 & 92102 & 1.34 \\
CXOMP~J141556.8+230727 & 2024 & 14 15 56.8 & +23 07 27.1 & 14755 & 1.91 \\
CXOMP~J141602.1+230647 & 2024 & 14 16 02.1 & +23 06 47.8 & 14755 & 1.91 \\
CXOMP~J153259.2$-$004414 & 2085 & 15 32 59.2 & $-$00 44 14.7 & 5152 & 6.25 \\
CXOMP~J153415.0+232459 & 869 & 15 34 15.0 & +23 24 59.7 & 57181 & 4.28 \\
CXOMP~J154932.0+213300 & 326 & 15 49 32.0 & +21 33 00.7 & 42688 & 4.30 \\
CXOMP~J160847.1+654139 & 2127 & 16 08 47.1 & +65 41 39.2 & 44648 & 2.83 \\
CXOMP~J160948.4+660056 & 2127 & 16 09 48.4 & +66 00 56.9 & 44648 & 2.83 \\
CXOMP~J165514.4$-$082944 & 615 & 16 55 14.4 & $-$08 29 44.0 & 9152 & 13.40 \\
CXOMP~J205537.4$-$043334 & 551 & 20 55 37.4 & $-$04 33 34.8 & 44880 & 4.96 \\
CXOMP~J205617.2$-$044154 & 551 & 20 56 17.2 & $-$04 41 54.8 & 44880 & 4.96 \\
CXOMP~J220455.8$-$181524 & 2114 & 22 04 55.8 & $-$18 15 24.3 & 5146 & 2.79 \\
CXOMP~J221326.2$-$220532 & 1479 & 22 13 26.2 & $-$22 05 32.4 & 20774 & 2.49 \\
CXOMP~J223538.4+340609 & 789 & 22 35 38.4 & +34 06 09.3 & 19955 & 7.74 \\
CXOMP~J223614.5+335648 & 789 & 22 36 14.5 & +33 56 48.4 & 19955 & 7.74 \\
CXOMP~J230150.7+084352 & 918 & 23 01 50.7 & +08 43 52.5 & 109955 & 5.05 \\
CXOMP~J230227.7+083901 & 918 & 23 02 27.7 & +08 39 01.4 & 109955 & 5.05 \\
CXOMP~J230252.0+084137 & 918 & 23 02 52.0 & +08 41 37.0 & 109955 & 5.05 \\
CXOMP~J230311.1+085131 & 918 & 23 03 11.1 & +08 51 31.2 & 109955 & 5.05 \\
CXOMP~J234817.8+010617 & 861 & 23 48 17.8 & +01 06 17.2 & 37322 & 3.81 \\
\enddata
\tablenotetext{a}{Vignetting-corrected exposure time.}
\tablenotetext{b}{Galactic $\mbox{N}_{\mbox{H}}$ values are taken from 
\citet{Stark92}.}
\end{deluxetable}

\begin{deluxetable}{cccccccccc}
\tablecaption{OPTICAL MOSAIC FIELDS}
\tabletypesize{\scriptsize}
\tablewidth{0pt}
\tablehead{
\colhead{ObsID} &\colhead{E($B-V$)\tablenotemark{a}} &\colhead{Telescope} &\colhead{UT Date} 
&\colhead{Filter} &\colhead{Dithers} &\colhead{Total Exp.} &\colhead{Air Mass} 
&\colhead{FWHM\tablenotemark{b}} &\colhead{$\mbox{M}_{\mbox{TO}}$\tablenotemark{c}}\\ 
&&&&&&\colhead{(sec)} &\colhead{(mean)} &\colhead{(\arcsec)} &\colhead{(mag)}\\
}
\startdata
326 & 0.046 & KPNO 4m & 2001 Jun. 13 & $g^{\prime}$ & 3 & 2100 & 1.05 & 1.1 & 24.88 \\
&&&& $r^{\prime}$ & 5 & 2000 & 1.15 & 1.3 & 24.38 \\
&&& 2001 Jun. 12 & $i^{\prime}$ & 15 & 4500 & 1.18 & 1.1 & 24.62 \\
363 & 0.041 & KPNO 4m & 2001 Jun. 13 & $g^{\prime}$ & 2 & 1200 & 2.26 & 1.8 & 23.62 \\
&&&& $r^{\prime}$ & 5 & 1500 & 1.94 & 1.6 & 23.88 \\
&&&& $i^{\prime}$ & 5 & 1500 & 1.46 & 1.3 & 23.12 \\
367 & 0.047 & KPNO 4m & 2004 Jun. 19 & $g^{\prime}$ & 3 & 900 & 1.50 & 1.3 & 24.38 \\
&&&& $r^{\prime}$ & 3 & 600 & 1.40 & 1.7 & 23.62 \\
&&&& $i^{\prime}$ & 3 & 600 & 1.32 & 1.1 & 23.88 \\
431 & 0.071 & KPNO 4m & 2000 Jun. 11 & $g^{\prime}$ & 2 & 1000 & 1.36 & 1.6 & 24.12 \\
&&&& $r^{\prime}$ & 1 & 500 & 1.28 & 1.6 & 23.38 \\
&&&& $i^{\prime}$ & 1 & 360 & 1.25 & 1.2 & 22.88 \\
507 & 0.061 & CTIO 4m & 2003 Apr. 7 & $g^{\prime}$ & 3 & 900 & 1.06 & 1.0 & 24.88 \\
&&&& $r^{\prime}$ & 3 & 600 & 1.08 & 1.0 & 24.38 \\
&&&& $i^{\prime}$ & 3 & 600 & 1.09 & 1.1 & 23.62 \\
512 & 0.034 & KPNO 4m & 2001 Feb. 22 & $g^{\prime}$ & 5 & 4500 & 1.24 & 1.3 & 24.88 \\
&&&& $r^{\prime}$ & 3 & 2400 & 1.24 & 1.1 & 24.38 \\
&&&& $i^{\prime}$ & 5 & 2000 & 1.32 & 1.3 & 23.62 \\
521 & 0.061 & KPNO 4m & 2001 Oct. 24 & $g^{\prime}$ & 5 & 3000 & 1.15 & 1.5 & 24.62 \\
&&&& $r^{\prime}$ & 5 & 3000 & 1.04 & 1.2 & 24.38 \\
&&&& $i^{\prime}$ & 5 & 3000 & 1.00 & 1.1 & 23.88 \\
541 & 0.007 & KPNO 4m & 2000 Jun. 12 & $g^{\prime}$ & 2 & 1000 & 1.33 & 1.9 & 22.88 \\
&&&& $r^{\prime}$ & 1 & 500 & 1.41 & 2.0 & 22.12 \\
&&&& $i^{\prime}$ & 1 & 500 & 1.50 & 1.7 & 22.12 \\
546 & 0.035 & KPNO 4m & 2000 Jun. 11 & $g^{\prime}$ & 2 & 1400 & 1.04 & 1.5 & 23.62 \\	
&&&& $r^{\prime}$ & 1 & 500 & 1.02 & 1.5 & 22.88 \\
&&&& $i^{\prime}$ & 1 & 500 & 1.02 & 1.2 & 23.12 \\
551 & 0.079 & KPNO 4m & 2000 Oct. 17 & $g^{\prime}$ & 2 & 1800 & 1.27 & 1.4 & 24.88 \\
&&&& $r^{\prime}$ & 3 & 1440 & 1.40 & 1.1 & 24.62 \\
&&&& $i^{\prime}$ & 3 & 1260 & 1.52 & 1.4 & 23.62 \\
796 & 0.046 & KPNO 4m & 2001 Oct. 24 & $g^{\prime}$ & 3 & 2700 & 1.38 & 1.1 & 25.12 \\
&&&& $r^{\prime}$ & 3 & 2400 & 1.61 & 1.2 & 24.62 \\
&&&& $i^{\prime}$ & 3 & 1200 & 1.30 & 1.1 & 23.62 \\
800 & 0.019 & KPNO 4m & 2001 Jun. 14 & $g^{\prime}$ & 3 & 2400 & 1.11 & 1.8 & 24.62 \\
&&&& $r^{\prime}$ & 3 & 2100 & 1.28 & 1.6 & 23.88 \\
&&&& $i^{\prime}$ & 7 & 2520 & 1.48 & 1.3 & 23.62 \\
813 & 0.015 & CTIO 4m & 2000 Sep. 30 & $g^{\prime}$ & 3 & 180 & 1.07 & 1.4 & 23.88 \\
&&&& $r^{\prime}$ & 2 & 120 & 1.09 & 1.3 & 23.62 \\
&&&& $i^{\prime}$ & 3 & 180 & 1.11 & 1.2 & 22.88 \\
842 & 0.058 & CTIO 4m & 2000 Sep. 30 & $g^{\prime}$ & 3 & 180 & 1.06 & 1.1 & 23.62 \\
&&&& $r^{\prime}$ & 3 & 180 & 1.06 & 1.0 & 23.62 \\
&&&& $i^{\prime}$ & 3 & 180 & 1.06 & 1.0 & 22.62 \\
861 & 0.025 & CTIO 4m & 2000 Sep. 29 & $g^{\prime}$ & 3 & 1260 & 1.24 & 1.4 & 24.62 \\
&&&& $r^{\prime}$ & 3 & 1080 & 1.17 & 1.4 & 24.38 \\
&&&& $i^{\prime}$ & 3 & 1170 & 1.17 & 1.1 & 23.62 \\
898 & 0.038 & CTIO 4m & 2003 Apr. 7 & $g^{\prime}$ & 3 & 1800 & 1.02 & 1.0 & 25.38 \\
&&&& $r^{\prime}$ & 3 & 1200 & 1.06 & 1.0 & 24.62 \\
&&&& $i^{\prime}$ & 3 & 900 & 1.10 & 0.8 & 23.88 \\
913 & 0.014 & KPNO 4m & 2001 Oct. 23 & $g^{\prime}$ & 5 & 3500 & 1.46 & 1.1 & 25.12 \\
&&&& $r^{\prime}$ & 5 & 3000 & 1.44 & 1.0 & 24.62 \\
&&&& $i^{\prime}$ & 5 & 2000 & 1.52 & 1.1 & 23.62 \\
914 & 0.036 & CTIO 4m & 2000 Sep. 29 & $g^{\prime}$ & 3 & 990 & 1.04 & 1.2 & 25.12 \\
&&&& $r^{\prime}$ & 3 & 810 & 1.06 & 1.2 & 24.38 \\
&&&& $i^{\prime}$ & 3 & 900 & 1.08 & 1.2 & 23.62 \\
915 & 0.051 & CTIO 4m & 2003 Apr. 7 & $g^{\prime}$ & 3 & 1800 & 1.02 & 1.0 & 25.38 \\
&&&& $r^{\prime}$ & 3 & 1200 & 1.00 & 1.0 & 24.62 \\
&&&& $i^{\prime}$ & 3 & 900 & 1.00 & 0.8 & 23.88 \\
918 & 0.081 & KPNO 4m & 2001 Oct. 23 & $g^{\prime}$ & 5 & 4500 & 1.13 & 1.2 & 24.88 \\
&&&& $r^{\prime}$ & 5 & 3000 & 1.09 & 1.3 & 24.38 \\
&&&& $i^{\prime}$ & 5 & 1500 & 1.14 & 1.2 & 23.38 \\
926 & 0.071 & CTIO 4m & 2003 Apr. 6 & $g^{\prime}$ & 3 & 1800 & 1.15 & 1.1 & 24.88 \\
&&& 2003 Apr. 7 & $r^{\prime}$ & 3 & 1200 & 1.10 & 1.2 & 24.38 \\
&&& 2003 Apr. 7 & $i^{\prime}$ & 3 & 900 & 1.07 & 0.9 & 23.88 \\
928 & 0.052 & CTIO 4m & 2000 Sep. 29 & $g^{\prime}$ & 3 & 900 & 1.01 & 1.6 & 24.38 \\
&&&& $r^{\prime}$ & 3 & 720 & 1.02 & 1.3 & 24.12 \\
&&&& $i^{\prime}$ & 3 & 810 & 1.04 & 1.0 & 23.62 \\
930 & 0.021 & KPNO 4m & 2001 Feb. 22 & $g^{\prime}$ & 3 & 2700 & 1.11 & 1.2 & 25.62 \\
&&&& $r^{\prime}$ & 3 & 2550 & 1.07 & 1.3 & 24.62 \\
&&&& $i^{\prime}$ & 4 & 2080 & 1.08 & 1.2 & 24.38 \\
1479 & 0.033 & CTIO 4m & 2001 Aug. 22 & $g^{\prime}$ & 2 & 1200 & 1.21 & 1.0 & 24.88 \\
&&&& $r^{\prime}$ & 2 & 900 & 1.39 & 1.0 & 24.38 \\
&&&& $i^{\prime}$ & 2 & 720 & 1.44 & 1.1 & 23.62 \\
1602 & 0.080 & KPNO 4m & 2001 Oct. 23 & $g^{\prime}$ & 1 & 800 & 1.56 & 1.4 & 24.38 \\
&&&& $r^{\prime}$ & 1 & 750 & 1.56 & 1.1 & 23.88 \\
&&&& $i^{\prime}$ & 1 & 400 & 1.56 & 1.0 & 23.12 \\
1644 & 0.030 & CTIO 4m & 2001 Aug. 9 & $g^{\prime}$ & 5 & 1805 & 1.08 & 1.0 & 24.38 \\
&&&& $r^{\prime}$ & 3 & 810 & 1.09 & 1.0 & 23.88 \\
&&&& $i^{\prime}$ & 3 & 570 & 1.10 & 0.9 & 23.62 \\
1657 & 0.027 & KPNO 4m & 2004 Jun. 17 & $g^{\prime}$ & 3 & 1800 & 1.08 & 1.1 & 24.88 \\
&&&& $r^{\prime}$ & 3 & 1200 & 1.16 & 1.0 & 24.38 \\
&&&& $i^{\prime}$ & 3 & 900 & 1.24 & 1.2 & 23.62 \\
1694 & 0.065 & KPNO 4m & 2004 Jun. 17 & $g^{\prime}$ & 3 & 1800 & 1.54 & 1.1 & 24.62 \\
&&&& $r^{\prime}$ & 3 & 1200 & 1.36 & 1.1 & 24.38 \\
&&&& $i^{\prime}$ & 3 & 900 & 1.28 & 0.9 & 23.62 \\
1899 & 0.041 & KPNO 4m & 2004 Jun. 17 & $g^{\prime}$ & 3 & 1800 & 1.03 & 1.0 & 25.12 \\
&&&& $r^{\prime}$ & 3 & 1200 & 1.03 & 1.0 & 24.62 \\
&&&& $i^{\prime}$ & 3 & 900 &  1.04 & 0.8 & 24.12 \\
2024 & 0.024 & KPNO 4m & 2004 Jun. 19 & $g^{\prime}$ & 3 & 900 & 1.17 & 0.9 &  24.88 \\
&&&& $r^{\prime}$ & 3 & 600 & 1.24 & 1.0 & 24.38 \\
&&&& $i^{\prime}$ & 3 & 600 & 1.30 & 0.9 & 23.88 \\
2099 & 0.044 & KPNO 4m & 2001 Dec. 14 & $g^{\prime}$ & 1 & 100 & 1.21 & 2.1 & 22.88 \\	
&&&& $r^{\prime}$ & 1 & 90 & 1.21 & 1.6 & 22.38 \\
&&&& $i^{\prime}$ & 1 & 85 & 1.21 & 1.3 & 21.38 \\
2113 & 0.026 &	CTIO 4m & 2001 Aug. 9 & $g^{\prime}$ & 1 & 400 & 1.06 & 1.0 & 23.62 \\
&&&& $r^{\prime}$ & 1 & 150 & 1.07 & 0.9 & 23.12 \\
&&&& $i^{\prime}$ & 1 & 120 & 1.08 & 0.9 & 22.12 \\
2114 & 0.030 & CTIO 4m & 2001 Aug. 9 & $g^{\prime}$ & 1 & 400 & 1.30 & 1.1 &  23.12 \\
&&&& $r^{\prime}$ & 1 & 150 & 1.34 & 1.1 & 22.88 \\
&&&& $i^{\prime}$ & 1 & 120 & 1.28 & 0.9 & 22.12 \\
2127 & 0.034 & KPNO 4m & 2004 Jun. 19 & $g^{\prime}$ & 3 & 1800 & 1.36 & 1.4 & 24.88 \\
&&&& $r^{\prime}$ & 3 & 1200 & 1.43 & 1.1 & 24.62 \\
&&&& $i^{\prime}$ & 3 & 900 & 1.54 & 1.4 & 23.62 \\
2210 & 0.014 & KPNO 4m & 2004 Jun. 18 & $g^{\prime}$ & 3 & 1800 & 1.15 & 1.7 & 24.38 \\
&&&& $r^{\prime}$ & 3 & 1200 & 1.10 & 1.4 & 24.12 \\
&&&& $i^{\prime}$ & 3 & 900  & 1.07 & 1.2 & 23.62 \\
2221 & 0.020 & KPNO 4m & 2004 Jun. 18 & $g^{\prime}$ & 3 & 1800 & 1.20 & 1.4 & 25.12 \\
&&&& $r^{\prime}$ & 3 & 1200 & 1.15 & 1.4 & 24.38 \\
&&&& $i^{\prime}$ & 3 & 900  & 1.12 & 1.2 & 23.88 \\
2228 & 0.009 & KPNO 4m & 2004 Jun. 19 & $g^{\prime}$ & 3 & 1800 & 1.07 & 0.9 & 25.12 \\
&&&& $r^{\prime}$ & 3 & 1200 & 1.13 & 0.8 & 24.88 \\
&&&& $i^{\prime}$ & 3 & 900 &  1.20 & 0.9 & 24.12 \\
\enddata
\tablenotetext{a}{Galactic extinction values are calculated from the maps 
of \citet{Schlegel98}.}
\tablenotetext{b}{FWHM of point sources in final stacked image.}
\tablenotetext{c}{Turnover magnitude of galaxy counts using 0.25 mag bins 
prior to extinction correction.}
\end{deluxetable}

\begin{deluxetable}{cccccccccccr}
\tablecaption{X-RAY PROPERTIES OF EXTENDED SOURCES}
\tabletypesize{\scriptsize}
\rotate
\tablewidth{0pt}
\tablehead{\colhead{Source Name} &\colhead{Counts} &\colhead{$f_X$} 
&\colhead{$\Delta f_X$\tablenotemark{a}}&\colhead{$L_X$\tablenotemark{b}} &\colhead{$\Delta L_X$}
&\colhead{$\mbox{r}_{\mbox{c}}$\tablenotemark{c}}&\colhead{PSF\tablenotemark{d}}&\colhead{OAA\tablenotemark{e}}
&\colhead{$z$\tablenotemark{f}}&\colhead{$z$\tablenotemark{g}}\\ 
&&\colhead{($10^{-14}$~cgs)} &\colhead{($10^{-14}$~cgs)}&\colhead{($10^{42}$~cgs)}
&\colhead{($10^{42}$~cgs)}&\colhead{(arcsec)}&\colhead{(arcsec)}&\colhead{(arcmin)}
&\colhead{(spec)} &\colhead{(VTP)}}
\startdata
CXOMP~J002650.2+171935 & 125.38 & 2.523 & 0.225 & 23.161 & 2.068 & 15.98 & 5.41 & 9.68 & 0.4907 & \nodata \\
CXOMP~J005848.1$-$280035 & 178.32 & 15.272 & 1.144 & 17.967 & 1.346 & 24.90 & 7.65 & 11.52 & 0.2021 & \nodata \\
CXOMP~J010214.1+314915 & 124.91 & 1.593 & 0.142 & 2.429 & 0.217 & 14.03 & 0.63 & 2.57 &\nodata & 0.227 \\
CXOMP~J010607.0+004943 & 186.98 & 23.209 & 1.697 & 55.593 & 4.066 & 40.34 & 0.76 & 3.14 & 0.2767 & \nodata \\
CXOMP~J010610.3+005126 & 120.07 & 22.936 & 2.093 & 48.930 & 4.465 & 41.15 & 0.81 & 3.32 & 0.2630 & \nodata \\
CXOMP~J013642.6+204843 & 229.06 & 3.632 & 0.240 & \nodata & \nodata & 27.45 & 4.84 & 9.16 & \nodata & \nodata \\
CXOMP~J033639.4$-$045515 & 600.58 & 13.281 & 0.542 & \nodata & \nodata & 39.28 & 17.96 & 17.67 & \nodata & \nodata \\
CXOMP~J033722.6$-$045906 & 166.37 & 2.822 & 0.219 & 0.022 & 0.002 & 3.07 & 2.33  & 6.29 & 0.0185 & \nodata \\
CXOMP~J033755.1$-$050733 & 2770.02 & 39.048 & 0.742 & 12.687 & 0.241 & 55.46 & 2.05 & 5.88 & 0.1123 & 0.102 \\
CXOMP~J033757.8$-$050001 & 64.83 & 0.889 & 0.110 & 0.026 & 0.003 & 2.32 & 1.07 & 4.06 & 0.0357\tablenotemark{h}& \nodata \\
CXOMP~J040351.2$-$170823 & 407.08 & 55.282 & 2.740 & \nodata & \nodata & 81.79 & 1.93 & 5.69 & \nodata & \nodata \\
CXOMP~J054152.7$-$410702 & 245.74 & 3.877 & 0.247 & \nodata & \nodata & 9.96 & 10.39 & 13.44 & \nodata & \nodata \\
CXOMP~J054240.1$-$405503 & 273.41 & 3.582 & 0.217 & 61.071 & 3.693 & 21.37 & 1.09 & 4.10 & 0.6340 & 0.627 \\
CXOMP~J063057.7+820701 & 201.10 & 4.166 & 0.294 & 9.458 & 0.667 & 35.50 & 6.93 & 10.96 & 0.2703\tablenotemark{h} & 0.302 \\
CXOMP~J090634.4+340055 & 74.72 & 7.081 & 0.819 & 25.292 & 2.926 & 19.64 & 13.28 & 15.19 & 0.3290 & \nodata \\
CXOMP~J091008.4+541852 & 2427.33 & 15.172 & 0.308 & \nodata & \nodata & 1.24 & 1.35 & 4.68 & \nodata & \nodata \\
CXOMP~J091126.6+055012 & 175.64 & 2.902 & 0.219 & 79.135 & 5.971 & 13.34 & 0.50 & 1.01 & 0.7682 & \nodata \\
CXOMP~J091301.4+054814 & 254.94 & 8.056 & 0.504 & \nodata & \nodata & 39.71 & 31.32 & 23.34 & \nodata & \nodata \\
CXOMP~J093102.2+791320 & 1384.00 & 43.783 & 1.177 & 384.842 & 10.345 & 30.19 & 3.32 & 7.56 & 0.4819\tablenotemark{h}& \nodata \\
CXOMP~J093352.9+552619 & 132.45 & 2.415 & 0.210 & \nodata & \nodata & 14.62 & 7.87 & 11.69 & \nodata & \nodata \\
CXOMP~J095012.8+142351 & 164.93 & 5.466 & 0.426 & \nodata & \nodata & 17.30 & 1.26 & 4.48 & \nodata & \nodata \\
CXOMP~J101008.7$-$124013 & 71.75 & 1.100 & 0.130 & \nodata & \nodata & 15.47 & 0.52 & 1.68 & \nodata & \nodata \\
CXOMP~J101115.3$-$124147 & 373.72 & 6.929 & 0.358 & \nodata & \nodata & 16.35 & 12.62 & 14.81 & \nodata & \nodata \\
CXOMP~J105624.6$-$033517 & 705.80 & 6.058 & 0.228 & 100.132 & 3.769 & 36.23 & 4.44 & 8.76 & 0.6260 & 0.602 \\
CXOMP~J111405.8+403157 & 61.29 & 1.559 & 0.199 & \nodata & \nodata & 23.12 & 2.88 & 7.03 & \nodata & \nodata \\
CXOMP~J111726.1+074335 & 286.03 & 9.866 & 0.583 & 84.623 & 5.004 & 32.88 & 8.70 & 12.29 & 0.4770 & 0.552 \\
CXOMP~J111730.2+074618 & 165.91 & 5.397 & 0.419 & 3.783 & 0.294 & 34.48 & 7.23 & 11.20 & 0.1600 & 0.177 \\
CXOMP~J114008.2$-$263132 & 233.82 & 5.119 & 0.335 & 111.844 & 7.314 & 33.85 & 4.23 & 8.55 & \nodata & 0.702 \\
CXOMP~J114118.8+660209 & 93.05 & 0.479 & 0.050 & \nodata & \nodata & 2.91 & 1.96 & 5.75 & \nodata & \nodata \\
CXOMP~J122927.1+752037 & 878.57 & 9.168 & 0.309 & \nodata & \nodata & 52.98 & 1.70 & 5.32 & \nodata & \nodata \\
CXOMP~J122940.6+752106 & 815.18 & 8.592 & 0.301 & \nodata & \nodata & 23.87 & 1.61 & 5.16 & \nodata & \nodata \\
CXOMP~J131709.9+285513 & 301.90 & 2.347 & 0.135 & \nodata & \nodata & 16.61 & 13.11 & 15.10 & \nodata & \nodata \\
CXOMP~J131722.0+285353 & 586.09 & 4.836 & 0.200 & 0.435 & 0.018 & 25.27 & 15.80 & 16.57 & 0.0612 & \nodata \\
CXOMP~J134507.8+000359 & 43.10 & 3.733 & 0.568 & 16.615 & 2.531 & 12.97 & 7.35 & 11.29 & 0.3616 & \nodata \\
CXOMP~J134514.6$-$000846 & 292.62 & 25.966 & 1.518 & 9.364 & 0.547 & 62.68 & 12.21 & 14.56 & 0.1179 & \nodata \\
CXOMP~J141152.6+520937 & 62.79 & 0.504 & 0.064 & \nodata & \nodata & 9.78 & 2.04 & 5.87 & \nodata & \nodata \\
CXOMP~J141556.8+230727 & 160.64 & 8.851 & 0.698 & 31.138 & 2.457 & 11.81 & 5.46 & 9.72 & \nodata & 0.327 \\
CXOMP~J141602.1+230647 & 94.96 & 5.211 & 0.535 & 34.307 & 3.520 & 33.59 & 5.07 & 9.37 & \nodata & 0.427 \\
CXOMP~J153259.2$-$004414 & 708.42 & 69.601 & 2.615 & \nodata & \nodata & 87.89 & 1.61 & 5.16 & \nodata & \nodata \\
CXOMP~J153415.0+232459 & 150.02 & 1.906 & 0.156 & \nodata & \nodata & 6.63 & 6.14 & 10.32 & \nodata & \nodata \\
CXOMP~J154932.0+213300 & 335.87 & 5.945 & 0.324 & 80.582 & 4.397 & 11.71 & 4.06 & 8.37 & \nodata & 0.577 \\
CXOMP~J160847.1+654139 & 202.61 & 4.912 & 0.345 & \nodata & \nodata & 26.57 & 14.22 & 15.72 & \nodata & \nodata \\
CXOMP~J160948.4+660056 & 378.22 & 4.414 & 0.227 & 48.071 & 2.472 & 22.04 & 1.32 & 4.61 & \nodata & 0.527 \\
CXOMP~J165514.4$-$082944 & 47.76 & 4.039 & 0.584 & \nodata & \nodata & 22.08 & 3.15 & 7.35 & \nodata & \nodata \\
CXOMP~J205537.4$-$043334 & 237.63 & 4.098 & 0.266 & \nodata & \nodata & 7.28 & 6.13 & 10.31 & \nodata & \nodata \\
CXOMP~J205617.2$-$044154 & 88.51 & 1.464 & 0.156 & 21.846 & 2.322 & 3.09 & 3.18 & 7.39 & 0.6002\tablenotemark{h} & 0.702 \\
CXOMP~J220455.8$-$181524 & 199.35 & 34.218 & 2.424 & \nodata & \nodata & 50.83 & 10.16 & 13.29 & \nodata & \nodata \\
CXOMP~J221326.2$-$220532 & 72.22 & 2.290 & 0.269 & 0.017 & 0.002 & 9.54 & 2.44 & 6.45 & 0.0180\tablenotemark{h} & \nodata \\
CXOMP~J223538.4+340609 & 149.88 & 5.406 & 0.442 & 26.510 & 2.165 & 3.58 & 3.52 & 7.78 & 0.3768\tablenotemark{h}& \nodata \\
CXOMP~J223614.5+335648 & 117.73 & 2.671 & 0.246 & \nodata & \nodata & 15.15 & 1.19 & 4.33 & \nodata & \nodata \\
CXOMP~J230150.7+084352 & 321.47 & 2.716 & 0.151 & 16.420 & 0.916 & 22.85 & 11.40 & 14.07 & 0.4118\tablenotemark{h}& \nodata \\
CXOMP~J230227.7+083901 & 97.82 & 0.677 & 0.068 & 4.427 & 0.448 & 7.85 & 3.62 & 7.90 & 0.4256\tablenotemark{h}& \nodata \\
CXOMP~J230252.0+084137 & 456.11 & 2.874 & 0.134 & 0.116 & 0.005 & 6.44 & 0.98 & 3.82 & 0.0415\tablenotemark{h}& \nodata \\
CXOMP~J230311.1+085131 & 685.24 & 4.813 & 0.184 & \nodata & \nodata & 38.75 & 4.27 & 8.58 & \nodata & \nodata \\
CXOMP~J234817.8+010617 & 384.19 & 4.844 & 0.247 & 1.057 & 0.054 & 15.62 & 3.42 & 7.68 & 0.0932\tablenotemark{h}& \nodata \\
\enddata
\tablenotetext{a}{All tabulated uncertainties are $1\sigma$ values.}
\tablenotetext{b}{Luminosities are calculated from spectroscopic redshifts 
unless only VTP estimates are available.}
\tablenotetext{c}{Source core radius is estimated from the circular 
$\beta$-model fits. The uncertainty in $\Delta r_{c}$ is on the order of 
10-20\%.} 
\tablenotetext{d}{{\it Chandra} PSF size at the location of extended X-ray 
source. The PSF is derived from the best-fit analytic relation between 
the Gaussian sigma of point sources, and is similar to the radius encircling 
50\% of total counts for a monochromatic source at 0.75 keV.}
\tablenotetext{e}{Off-axis angle.}
\tablenotetext{f}{Redshifts obtained from the literature have NED IDs 
listed in Table~5. Those marked by a superscript ``h'' are from ChaMP 
spectroscopy. Typical spectroscopic redshift uncertainties are $\sim 0.0006$.} 
\tablenotetext{g}{Red-sequence filtered VTP corrected redshifts. The dispersion of 
the VTP redshifts about the spectroscopic values is 0.03.}
\tablenotetext{h}{Redshifts measured from our ChaMP spectroscopic program.}
\end{deluxetable}

\clearpage

\begin{deluxetable}{cccccclc}
\tablecaption{VTP CLUSTERS WITH X-RAY UPPER LIMITS}
\tabletypesize{\small}
\tablewidth{0pt}
\tablehead{\colhead{No.} &\colhead{ObsID}&\colhead{RA}&\colhead{DEC}&\colhead{Redshift\tablenotemark{a}}
&\colhead{$f_X\tablenotemark{b}$}&\colhead{$L_X$}&\colhead{PSF\tablenotemark{c}}\\
&&\colhead{(J2000)}&\colhead{(J2000)}&&\colhead{($10^{-14}$~cgs)}&\colhead{($10^{42}$~cgs)}
&\colhead{(arcsec)}}
\startdata
1 & 2099 & 00 23 59.1 & $-$01 50 17.0 & 0.127 & $\leq 2.947$ & $\leq 1.256$ & 2.52\\
2 & 521 & 01 01 27.6 & +31 46 46.9 & 0.227 & $\leq 1.167$ & $\leq 1.795$ & 3.47\\
3 & 521 & 01 01 31.0 & +31 46 44.4 & 0.402 & $\leq 1.207$ & $\leq 6.942$ & 2.88\\
4 & 521 & 01 01 54.7 & +31 45 38.1 & 0.502 & $\leq 1.181$ & $\leq 11.518$ & 0.73\\
5 & 521 & 01 02 08.7 & +31 55 55.0 & 0.577 & $\leq 1.166$ & $\leq 15.901$ & 3.67\\
6 & 813 & 01 02 53.8 & $-$27 07 22.8 & 0.221\tablenotemark{d} & $\leq 3.664$ & $\leq 5.321$ & 12.37\\
7 & 913 & 01 52 44.9 & $-$14 01 32.6 & 0.527 & $\leq 1.244$ & $\leq 13.640$ & 1.72\\
8 & 913 & 01 53 16.0 & $-$13 57 18.0 & 0.577 & $\leq 1.269$ & $\leq 17.309$ & 2.53\\
9 & 913 & 01 53 18.1 & $-$13 52 00.4 & 0.177 & $\leq 1.303$ & $\leq 1.149$ & 3.91\\
10 & 796 & 03 36 41.7 & $-$04 53 52.0 & 0.352 & $\leq 2.022$ & $\leq 8.514$ & 18.13\\
11 & 796 & 03 36 42.4 & $-$04 59 30.4 & 0.727 & $\leq 2.238$ & $\leq 53.553$ & 14.19\\
12 & 796 & 03 37 31.8 & $-$05 10 21.7 & 0.202 & $\leq 1.070$ & $\leq 1.266$ & 4.32\\
13 & 796 & 03 38 01.7 & $-$04 53 43.9 & 0.527 & $\leq 1.133$ & $\leq 12.420$ & 5.32\\
14 & 914 & 05 41 38.7 & $-$41 10 09.2 & 0.527 & $\leq 1.142$ & $\leq 12.524$ & 17.50\\ 
15 & 914 & 05 42 24.9 & $-$41 00 14.2 & 0.502 & $\leq 1.114$ & $\leq 10.866$ & 1.40\\
16 & 914 & 05 42 25.1 & $-$40 53 12.9 & 0.577 & $\leq 0.914$ & $\leq 12.469$ & 2.99\\
17 & 914 & 05 42 36.3 & $-$40 50 07.6 & 0.527 & $\leq 0.939$ & $\leq 10.296$ & 4.67\\
18 & 914 & 05 42 57.5 & $-$40 58 08.2 & 0.527 & $\leq 0.930$ & $\leq 10.200$ & 0.53\\
19 & 1602 & 06 24 59.1 & +81 59 09.6 & 0.252 & $\leq 0.981$ & $\leq 1.911$ & 1.23\\
20 & 926 & 10 09 28.8 & $-$12 44 34.8 & 0.402 & $\leq 0.999$ & $\leq 5.746$ & 7.92\\ 
21 & 926 & 10 09 35.8 & $-$12 43 54.4 & 0.602 & $\leq 1.344$ & $\leq 20.314$ & 5.66\\
22 & 926 & 10 09 39.9 & $-$12 45 54.6 & 0.277 & $\leq 1.219$ & $\leq 2.947$ & 5.50\\
23 & 926 & 10 09 48.1 & $-$12 38 49.5 & 0.352 & $\leq 1.222$ & $\leq 5.145$ & 2.75\\
24 & 926 & 10 11 05.7 & $-$12 40 30.5 & 0.277 & $\leq 1.446$ & $\leq 3.497$ & 8.94\\
25 & 512 & 10 56 48.9 & $-$03 37 25.5 & 0.182\tablenotemark{e} & $\leq 0.902$ & $\leq 0.846$ & 0.63\\
26 & 512 & 10 57 00.5 & $-$03 44 19.6 & 0.277 & $\leq 0.692$ & $\leq 1.674$ & 1.60\\
27 & 915 & 11 13 44.2 & $-$26 23 38.4 & 0.627 & $\leq 1.616$ & $\leq 26.959$ & 10.91\\
28 & 363 & 11 17 20.4 & +07 58 56.3 & 0.577 & $\leq 1.961$ & $\leq 26.753$ & 21.31\\
29 & 363 & 11 17 35.7 & +07 42 52.1 & 0.527 & $\leq 1.490$ & $\leq 16.330$ & 5.86\\
30 & 363 & 11 17 41.8 & +07 45 03.6 & 0.402 & $\leq 1.348$ & $\leq 7.753$ & 3.98\\
31 & 363 & 11 18 16.6 & +07 43 23.9 & 0.277 & $\leq 2.550$ & $\leq 6.168$ & 0.54\\
32 & 898 & 11 39 50.4 & $-$26 34 23.3 & 0.702 & $\leq 2.043$ & $\leq 44.869$ & 9.85\\
33 & 898 & 11 40 40.4 & $-$26 34 02.8 & 0.577 & $\leq 1.640$ & $\leq 22.368$ & 1.01\\
34 & 898 & 11 40 46.8 & $-$26 34 44.4 & 0.452 & $\leq 1.136$ & $\leq 8.625$ & 1.22\\
35 & 898 & 11 40 52.2 & $-$26 24 07.3 & 0.427 & $\leq 2.293$ & $\leq 15.208$ & 2.38\\
36 & 2210 & 12 56 21.3 & +47 15 55.7 & 0.209\tablenotemark{d} & $\leq 1.089$ & $\leq 1.390$ & 3.21\\
37 & 2210 & 12 56 44.1 & +47 18 43.6 & 0.577 & $\leq 1.696$ & $\leq 23.133$ & 0.65\\
38 & 2228 & 13 16 40.7 & +29 06 23.3 & 0.652 & $\leq 0.606$ & $\leq 11.110$ & 3.63\\
39 & 2228 & 13 16 54.0 & +29 14 20.8 & 0.477 & $\leq 0.641$ & $\leq 5.532$ & 1.90\\
40 & 2228 & 13 17 21.1 & +29 20 42.0 & 0.652 & $\leq 0.630$ & $\leq 11.552$ & 6.44\\
41 & 2228 & 13 17 22.8 & +28 58 48.0 & 0.377 & $\leq 0.704$ & $\leq 3.483$ & 7.95\\
42 & 507 & 13 47 18.5 & $-$11 52 26.7 & 0.402 & $\leq 2.476$ & $\leq 14.241$ & 2.54\\
43 & 507 & 13 47 27.7 & $-$11 40 38.9 & 0.086\tablenotemark{d} & $\leq 3.428$ & $\leq 0.641$ & 1.95\\
44 & 2024 & 14 15 23.3 & +23 11 52.1 & 0.552 & $\leq 2.201$ & $\leq 26.976$ & 12.95\\
45 & 2024 & 14 15 44.1 & +23 14 25.4 & 0.252 & $\leq 2.534$ & $\leq 4.936$ & 5.84\\
46 & 2024 & 14 15 50.1 & +23 13 58.5 & 0.352 & $\leq 2.373$ & $\leq 9.990$ & 4.37\\
47 & 930 & 14 15 50.9 & +11 32 52.8 & 0.227 & $\leq 2.504$ & $\leq 3.850$ & 0.81\\
48 & 930 & 14 15 56.4 & +11 30 14.4 & 0.527 & $\leq 2.645$ & $\leq 29.000$ & 0.74\\
49 & 930 & 14 16 03.2 & +11 25 14.4 & 0.702 & $\leq 2.255$ & $\leq 49.541$ & 2.66\\
50 & 541 & 14 16 09.6 & +44 44 02.4 & 0.427 & $\leq 1.462$ & $\leq 9.699$ & 3.37\\
51 & 2024 & 14 16 19.2 & +23 05 59.3 & 0.577 & $\leq 1.806$ & $\leq 24.631$ & 3.99\\
52 & 2024 & 14 16 19.7 & +23 19 58.8 & 0.752 & $\leq 2.630$ & $\leq 68.346$ & 2.28\\
53 & 541 & 14 16 27.6 & +44 52 44.4 & 0.452 & $\leq 1.297$ & $\leq 9.851$ & 1.73\\
54 & 1657 & 14 22 56.9 & +24 08 27.6 & 0.227 & $\leq 1.828$ & $\leq 2.811$ & 6.46\\
55 & 1657 & 14 23 05.1 & +24 00 24.0 & 0.502 & $\leq 1.826$ & $\leq 17.816$ & 3.26\\
56 & 1657 & 14 23 35.0 & +23 49 48.2 & 0.502 & $\leq 2.128$ & $\leq 20.761$ & 9.76\\
57 & 367 & 14 23 55.9 & +23 03 36.0 & 0.427 & $\leq 1.787$ & $\leq 11.853$ & 8.35\\
58 & 367 & 14 24 37.0 & +23 05 56.4 & 0.227 & $\leq 1.494$ & $\leq 2.298$ & 6.01\\
59 & 367 & 14 24 43.0 & +23 06 14.4 & 0.477 & $\leq 0.964$ & $\leq 8.322$ & 6.52\\
60 & 367 & 14 25 04.6 & +22 56 20.4 & 0.627 & $\leq 2.315$ & $\leq 38.627$ & 2.63\\
61 & 800 & 15 13 43.2 & +36 43 55.2 & 0.202 & $\leq 1.569$ & $\leq 1.856$ & 6.31\\
62 & 800 & 15 14 29.3 & +36 40 33.6 & 0.252 & $\leq 1.436$ & $\leq 2.798$ & 1.46\\
63 & 326 & 15 48 50.9 & +21 29 06.0 & 0.277 & $\leq 1.262$ & $\leq 3.052$ & 10.40\\
64 & 326 & 15 49 01.3 & +21 30 16.7 & 0.377 & $\leq 1.322$ & $\leq 6.534$ & 7.74\\
65 & 326 & 15 49 05.1 & +21 20 42.3 & 0.577 & $\leq 0.841$ & $\leq 11.474$ & 6.55\\
66 & 326 & 15 49 31.6 & +21 23 36.7 & 0.502 & $\leq 1.029$ & $\leq 10.038$ & 0.98\\
67 & 326 & 15 49 41.8 & +21 29 00.3 & 0.227 & $\leq 1.638$ & $\leq 2.519$ & 0.97\\
68 & 2127 & 16 08 10.1 & +65 44 20.4 & 0.302 & $\leq 0.984$ & $\leq 2.904$ & 13.24\\
69 & 2127 & 16 08 27.1 & +65 46 22.8 & 0.477 & $\leq 1.303$ & $\leq 11.250$ & 8.98\\
70 & 546 & 16 22 33.3 & +26 30 44.4 & 0.177 & $\leq 1.737$ & $\leq 1.532$ & 9.54\\
71 & 2221 & 17 14 08.4 & +50 20 56.4 & 0.627 & $\leq 0.932$ & $\leq 15.546$ & 2.08\\
72 & 2221 & 17 14 17.5 & +50 02 56.4 & 0.527 & $\leq 1.077$ & $\leq 11.811$ & 8.58\\
73 & 1899 & 18 06 25.9 & +45 54 52.3 & 0.627 & $\leq 1.253$ & $\leq 20.900$ & 7.25\\
74 & 1899 & 18 06 32.2 & +46 00 54.0 & 0.677 & $\leq 1.372$ & $\leq 27.576$ & 10.53\\
75 & 1899 & 18 07 17.5 & +45 47 02.4 & 0.252 & $\leq 0.978$ & $\leq 1.905$ & 1.30\\
76 & 1899 & 18 07 48.0 & +45 56 49.2 & 0.677 & $\leq 0.941$ & $\leq 18.923$ & 2.56\\
77 & 842 & 20 10 48.2 & $-$48 50 02.4 & 0.477 & $\leq 5.484$ & $\leq 47.342$ & 10.78\\
78 & 551 & 20 55 42.8 & $-$04 33 55.7 & 0.277 & $\leq 0.970$ & $\leq 2.346$ & 4.63\\
79 & 551 & 20 56 52.0 & $-$04 39 11.7 & 0.227 & $\leq 1.145$ & $\leq 1.760$ & 5.25\\
80 & 928 & 21 39 27.3 & $-$23 42 26.9 & 0.677 & $\leq 1.211$ & $\leq 24.347$ & 6.72\\
81 & 928 & 21 39 30.5 & $-$23 36 57.6 & 0.352 & $\leq 1.096$ & $\leq 4.613$ & 6.50\\
82 & 928 & 21 40 24.5 & $-$23 42 41.2 & 0.552 & $\leq 0.898$ & $\leq 11.013$ & 0.75\\
83 & 1644 & 21 51 16.7 & $-$27 34 50.9 & 0.327 & $\leq 2.830$ & $\leq 10.038$ & 7.70\\
84 & 2113 & 21 57 08.0 & $-$19 51 27.1 & 0.352 & $\leq 4.629$ & $\leq 19.488$ & 0.50\\
85 & 2114 & 22 04 52.8 & $-$18 15 36.2 & 0.277 & $\leq 8.549$ & $\leq 20.673$ & 9.08\\
86 & 1479 & 22 12 53.5 & $-$22 09 39.0 & 0.202 & $\leq 1.644$ & $\leq 1.945$ & 1.24\\
87 & 1479 & 22 12 55.7 & $-$22 12 10.5 & 0.652 & $\leq 1.602$ & $\leq 29.390$ & 1.00\\
88 & 1479 & 22 13 15.8 & $-$22 17 44.4 & 0.477 & $\leq 1.491$ & $\leq 12.869$ & 2.67\\
89 & 1479 & 22 13 30.2 & $-$22 03 33.8 & 0.677 & $\leq 1.531$ & $\leq 30.791$ & 4.30\\
90 & 1694 & 22 17 46.8 & +00 22 44.4 & 0.602 & $\leq 0.905$ & $\leq 13.686$ & 4.56\\
91 & 1694 & 22 17 50.6 & +00 21 32.4 & 0.227 & $\leq 0.873$ & $\leq 1.342$ & 4.17\\
92 & 1694 & 22 18 33.1 & +00 17 37.2 & 0.332\tablenotemark{d} & $\leq 0.743$ & $\leq 2.736$ & 15.48\\
93 & 431 & 22 39 55.2 & +03 38 26.8 & 0.127 & $\leq 1.996$ & $\leq 0.850$ & 20.65\\
94 & 431 & 22 39 56.5 & +03 31 33.7 & 0.302 & $\leq 1.577$ & $\leq 4.651$ & 9.54\\
95 & 431 & 22 40 25.0 & +03 31 15.6 & 0.427 & $\leq 1.662$ & $\leq 11.022$ & 5.72\\
96 & 431 & 22 40 25.4 & +03 33 56.2 & 0.252 & $\leq 1.755$ & $\leq 3.420$ & 9.17\\
97 & 918 & 23 02 51.8 & +08 50 20.5 & 0.402 & $\leq 0.672$ & $\leq 3.868$ & 1.63\\
98 & 918 & 23 03 09.4 & +08 49 19.1 & 0.502 & $\leq 0.708$ & $\leq 6.906$ & 2.70\\
99 & 861 & 23 47 59.3 & +01 03 44.5 & 0.248\tablenotemark{e} & $\leq 0.822$ & $\leq 1.549$ & 2.83\\
100 & 861 & 23 48 15.8 & +00 53 54.1 & 0.410\tablenotemark{e} & $\leq 0.899$ & $\leq 5.423$ & 1.39\\
101 & 861 & 23 48 39.6 & +01 08 40.9 & 0.527 & $\leq 1.010$ & $\leq 11.075$ & 7.52\\
102 & 861 & 23 49 35.6 & +00 59 40.6 & 0.352 & $\leq 1.125$ & $\leq 4.736$ & 21.68\\
\enddata
\tablenotetext{a}{Corrected redshifts estimated from red-sequence VTP unless 
otherwise noted.} 
\tablenotetext{b}{X-ray fluxes and luminosities are $3\sigma$ upper limits 
for cluster candidates not detected in X-rays.}
\tablenotetext{c}{{\it Chandra} PSF size at the location of optically detected source. 
The PSF is derived from the best-fit analytic relation between the Gaussian sigma 
of point sources, and is similar to the radius encircling 50\% of total counts for a 
monochromatic source at 0.75 keV.}
\tablenotetext{d}{Redshift measurement from NASA/IPAC Extragalactic Database.}
\tablenotetext{e}{Redshift measurement from ChaMP spectroscopic program.}
\end{deluxetable}

\begin{deluxetable}{cll}
\tablecaption{PUBLISHED SOURCES NEAR X-RAY POSITIONS
\tablenotemark{a}} 
\tabletypesize{\footnotesize}
\tablewidth{0pt}
\tablehead{\colhead{Source Name} &\colhead{Published ID\tablenotemark{b}} &\colhead{Comments\tablenotemark{c}}\\ }
\startdata
CXOMP~J002650.2+171935 & Zw Cl 0024.0+1652:[CKS2001] 541 & background group ($6\arcsec$)\tablenotemark{d}\\
CXOMP~J005848.1$-$280035 & 2MASX J00584850-2800414 & single galaxy:BCG? ($6\arcsec$) \\
CXOMP~J010214.1+314915 & 2MASX J01021352+3149243 & single galaxy:BCG? ($12\arcsec$)\\
CXOMP~J010607.0+004943 & SDSS CE J016.528793+00.817471 & poss. assoc. ($42\arcsec$) \\
CXOMP~J010610.3+005126 & SDSS J010610.38+005120.4 & galaxy ($6\arcsec$)\\
CXOMP~J013642.6+204843 & [B2002a] 02 & DSS: faint gals near X-ray pos. ($12\arcsec$)\\
CXOMP~J033639.4$-$045515 & \nodata  & \nodata \\
CXOMP~J033722.6$-$045906 & 2MASX J03372263-0459055 & single spiral galaxy ($0\arcsec$)\\
CXOMP~J033755.1$-$050733 & Abell 447 & galaxy cluster ($60\arcsec$)\\
CXOMP~J033757.8$-$050001 & 2MASX J03375780-0500006 & galaxy ($0\arcsec$)\\
CXOMP~J040351.2$-$170823 & APMUKS(BJ) B040135.74-171628.1 & galaxy ($6\arcsec$)\\
CXOMP~J054152.7$-$410702 & \nodata  & \nodata \\
CXOMP~J054240.1$-$405503 & RX J0542.8-4100 & galaxy cluster ($6\arcsec$)\\
CXOMP~J063057.7+820701 & 1WGA J0630.7+8206 & X-ray source ($30\arcsec$)\\
CXOMP~J090634.4+340055 & RIXOS F257\_037 &  AGN ($12\arcsec$)\\
CXOMP~J091008.4+541852 & CXOU J0910.1+5419: [B2002a] 14 & galaxy cluster ($0\arcsec$)\\
CXOMP~J091126.6+055012 & RX J0911.4+0551 & galaxy cluster ($0\arcsec$)\\
CXOMP~J091301.4+054814 & \nodata & \nodata \\
CXOMP~J093102.2+791320 & [B2002a] 15 & extended X-ray source ($6\arcsec$)\\
CXOMP~J093352.9+552619 & \nodata & \nodata \\
CXOMP~J095012.8+142351 & \nodata & \nodata \\
CXOMP~J101008.7$-$124013 & LCRS B100740.9-122545 & galaxy ($24\arcsec$) \\
CXOMP~J101115.3$-$124147 & \nodata & \nodata \\
CXOMP~J105624.6$-$033517 & XBS J105624.2-033522 & galaxy ($6\arcsec$) \\
CXOMP~J111405.8+403157 & \nodata & \nodata \\
CXOMP~J111726.1+074335 & RX J1117.4+0743 & galaxy cluster ($0\arcsec$) \\
CXOMP~J111730.2+074618 & RIXOS F258\_101 & galaxy cluster ($30\arcsec$) \\ 
CXOMP~J114008.2$-$263132 & [B2002a] 22 & extended X-ray src: gal. clus. ($0\arcsec$)\\
CXOMP~J114118.8+660209 & \nodata & \nodata \\
CXOMP~J122927.1+752037 & 1WGA J1229.6+7520 & X-ray source ($36\arcsec$) \\
CXOMP~J122940.6+752106 & 1WGA J1229.6+7520 & X-ray source ($30\arcsec$) \\
CXOMP~J131709.9+285513 & CXOSEXSI J131710.0+285516 & ($6\arcsec$) \\
CXOMP~J131722.0+285353 & 2MASX J13172206+2853460 & elliptical galaxy ($6\arcsec$) \\
CXOMP~J134507.8+000359 & 2QZ J134507.4+000406 & galaxy ($12\arcsec$) \\
CXOMP~J134514.6$-$000846 & [DDM2004] J134515.60-000830.8 & galaxy cluster ($24\arcsec$) \\
CXOMP~J141152.6+520937 & CXOSEXSI J141153.0+521020 & X-ray source ($42\arcsec$) \\ 
CXOMP~J141556.8+230727 & OC03 J1415+2307 & galaxy cluster ($12\arcsec$) \\
CXOMP~J141602.1+230647 & \nodata & \nodata \\
CXOMP~J153259.2$-$004414 & \nodata & \nodata \\
CXOMP~J153415.0+232459 & \nodata & \nodata \\ 
CXOMP~J154932.0+213300 & [B2002a] 30 & X-ray src: galaxy cluster ($36\arcsec$) \\
CXOMP~J160847.1+654139 & 2MASX J16084763+6541402 & galaxy ($6\arcsec$) \\
CXOMP~J160948.4+660056 & 1WGA J1609.7+6600 & X-ray source ($30\arcsec$) \\
CXOMP~J165514.4$-$082944 & \nodata & \nodata \\
CXOMP~J205537.4$-$043334 & CXOSEXSI J205537.3-043333 & X-ray source ($0\arcsec$) \\
CXOMP~J205617.2$-$044154 & CXOSEXSI J205617.1-044155 & X-ray source ($0\arcsec$) \\
CXOMP~J220455.8$-$181524 & \nodata & \nodata \\
CXOMP~J221326.2$-$220532 & IC 1435 & S-gal.: X-ray coincident with S-arm ($18\arcsec$) \\
CXOMP~J223538.4+340609 & 1WGA J2235.6+3406 & X-ray source ($12\arcsec$) \\
CXOMP~J223614.5+335648 & CXOU J223615.0+335630 & X-ray source ($18\arcsec$) \\
CXOMP~J230150.7+084352 & \nodata & \nodata \\
CXOMP~J230227.7+083901 & \nodata & \nodata \\
CXOMP~J230252.0+084137 & 2MASX J23025207+0841356 & single E-galaxy ($0\arcsec$) \\
CXOMP~J230311.1+085131 & \nodata & \nodata \\
CXOMP~J234817.8+010617 & 2MASX J23481801+0106174 & single E-galaxy ($0\arcsec$) \\
\enddata
\tablenotetext{a}{Published references acquired from the NASA/IPAC 
Extragalactic Database.}
\tablenotetext{b}{Blank field if published sources $>2\arcmin$ from object position.}
\tablenotetext{c}{Available redshifts tabulated in Table 3.}
\tablenotetext{d}{Distance in arcsec from extended X-ray centroid is shown in parentheses.} 
\end{deluxetable}

\clearpage

\begin{deluxetable}{cll}
\tablecaption{PUBLISHED SOURCES NEAR OPTICAL-ONLY DETECTIONS
\tablenotemark{a}} 
\tabletypesize{\footnotesize}
\tablewidth{0pt}
\tablehead{\colhead{No.} &\colhead{Published ID\tablenotemark{b}} &\colhead{Comments}\\ }
\startdata
1 & 2MASX J00235911-0150171 & galaxy: $0.0\arcsec$ from optical center\\ 
2 & 1WGA J0101.3+3146 & X-ray source within $60\arcsec$ of optical centroid \\
3 & \nodata & \nodata \\
4 & NVSS J010156+314537 & radio source within $24\arcsec$ of optical centroid \\
5 & \nodata & \nodata \\
6 & 2MASX J01025380-2707225 & galaxy ($z=0.221182$): $0.0\arcsec$ from opt. center \\ 
7 & RX J0152.7-1357 & galaxy: $54\arcsec$ from optical source \\ 
8 & CXOMP J015312.3-135723 & galaxy/ChaMP X-ray point src: $54\arcsec$ from opt. center \\
9 & CXOMP J015311.1-135104 & galaxy/ChaMP X-ray point src: $114\arcsec$ from opt. center \\
10 & \nodata & \nodata \\
11 & APMUKS(BJ) B033414.59-050957.3 & galaxy: $42\arcsec$ from optical source \\
12 & APMUKS(BJ) B033500.27-051957.8 & galaxy: $42\arcsec$ from optical source \\
13 & \nodata & \nodata \\
14 & \nodata & \nodata \\
15 & \nodata & \nodata \\
16 & \nodata & \nodata \\
17 & 2MASX J05422628-4049430 & galaxy: $114\arcsec$ from optical source \\ 
18 & \nodata & \nodata \\
19 & \nodata & \nodata \\
20 & \nodata & \nodata \\
21 & \nodata & \nodata \\
22 & \nodata & \nodata \\
23 & \nodata & \nodata \\
24 & \nodata & \nodata \\
25 & LCRS B105416.2-032123 & galaxy: $0.0\arcsec$ from optical centroid \\
26 & SPS J105700.03-034400.9 & galaxy: $18\arcsec$ from optical source position \\
27 & \nodata & \nodata \\
28 & NVSS J111718+075856 & radio source: $24\arcsec$ of optical cluster \\
29 & MAPS-NGP O\_553\_0096538 & galaxy: $96\arcsec$ from opt. source \\ 
30 & MAPS-NGP O\_553\_0096538 & galaxy: $66\arcsec$ from opt. source \\
31 & SDSS J111816.59+074323.9 & galaxy ($z=0.225281$): $0.0\arcsec$ from opt. src. \\
32 & NVSS J113953-263357 & radio source: $48\arcsec$ from optical source \\ 
33 & \nodata & \nodata \\
34 & \nodata & \nodata \\
35 & PKS 1138-26:[PKC2002] 08 & $102\arcsec$ from optical source \\ 
36 & SDSS J125621.26+471555.4 & galaxy ($z=0.208849$): $0.0\arcsec$ from opt. src.\\
37 & \nodata & \nodata \\
38 & CXOSEXSI J131637.2+290630 & X-ray source: $48\arcsec$ from optical cluster candidate \\
39 & CXOU J1316.9+2914 & galaxy cluster: $0.0\arcsec$ from opt. src.\\
40 & CXOSEXSI J131714.0+292034 & X-ray source: $90\arcsec$ from optical centroid \\
41 & CXOSEXSI J131721.8+285926 & X-ray source: $42\arcsec$ from optical source \\
42 & NVSS J134719-115226 & radio source within $12\arcsec$ of optical centroid \\
43 & 2MASX J13472770-1140397 & galaxy ($z=0.086372$): $0.0\arcsec$ from opt. center \\
44 & OC03a J1415+2311 & galaxy cluster: $48\arcsec$ from optical centroid ($z=0.500$ EST)\\
45 & 2MASX J14153929+2313477 & galaxy: $78\arcsec$ from optical source \\
46 & MAPS-NGP O\_382\_0330729 & galaxy: $90\arcsec$ from optical cluster candidate \\
47 & MAPS-NGP O\_500\_0168801 & galaxy: $54\arcsec$ from optical source \\
48 & [KSC90] 39 & galaxy: $120\arcsec$ from optical source \\
49 & NVSS J141601+112552 & radio source within $42\arcsec$ of optical centroid \\
50 & SDSS J141609.07+444416.6 & galaxy ($z=0.373182$): $18\arcsec$ from optical centroid \\ 
51 & 2MASX J14162498+2304411 & galaxy: $114\arcsec$ from optical cluster position \\ 
52 & 1WGA J1416.2+2321 & X-ray source: $114\arcsec$ from optical cluster position \\
53 & CXOMP J141626.6+445240 & galaxy/ChaMP X-ray point src: $12\arcsec$ from opt. center \\
54 & \nodata & \nodata \\
55 & \nodata & \nodata \\
56 & NVSS J142337+235135 & radio source within $114\arcsec$ of optical centroid \\
57 & \nodata & \nodata \\
58 & \nodata & \nodata \\ 
59 & 87GB 142237.3+231854 & radio source within $108\arcsec$ of optical centroid \\
60 & NVSS J142507+225642 & radio source within $48\arcsec$ of optical centroid \\
61 & SDSS J151346.73+364323.6 & galaxy ($z=0.240430$): $54\arcsec$ from optical centroid \\
62 & MS 1512.4+3647:PPP 102417 & galaxy ($z=403390$): $12\arcsec$ from optical centroid \\
63 & 2MASX J15484523+2130178 & galaxy: $108\arcsec$ from optical source \\
64 & NVSS J154901+213031 & radio source: $18\arcsec$ from optical centroid \\
65 & OC02a J1549+2119 & galaxy cluster: $66\arcsec$ from opt. src ($z=0.200$ EST)\\
66 & OC03 J1549+2123 & galaxy clusters: $12\arcsec$ from opt. center ($z=0.700$ EST)\\
67 & 3C 324 C006 & galaxy: $48\arcsec$ from optical source \\
68 & NVSS J160753+654404 & radio source: $108\arcsec$ from optical cluster position \\
69 & NVSS J160832+654437 & radio source: $108\arcsec$ from optical cluster position \\
70 & 2MASX J16223328+2630454 & galaxy: $0.0\arcsec$ from opt. centroid \\
71 & [KCW99] 28 & $102\arcsec$ from optical source \\
72 & 2MASX J17142140+5002487 & galaxy: $36\arcsec$ from optical source \\
73 & \nodata & \nodata \\
74 & \nodata & \nodata \\
75 & 2MASX J18072389+4546117 & galaxy: $84\arcsec$ from optical source \\
76 & \nodata & \nodata \\
77 & \nodata & \nodata \\
78 & CXOSEXSI J205537.3-043333 & X-ray source: $84\arcsec$ from optical source \\
79 & CXOSEXSI J205649.1-044013 & X-ray source: $72\arcsec$ from optical centroid \\
80 & CXOMP J213924.9-234221 & gal./ChaMP X-ray point src ($z=0.401$): $36\arcsec$ from opt. src \\
81 & \nodata & \nodata \\
82 & APMUKS(BJ) B213729.27-235708.1 & galaxy: $78\arcsec$ from optical centroid \\
83 & \nodata & \nodata \\
84 & LBQS 2154-2005 & QSO ($z=2.035000$): $30\arcsec$ from optical source \\
85 & 2MASX J22045283-1815366 & galaxy: $0.0\arcsec$ from opt. centroid \\
86 & [B2002a] 35 & galaxy cluster:  $0.0\arcsec$ from opt. src.\\
87 & APMUKS(BJ) B221007.01-222620.9 & $54\arcsec$ from optical centroid \\
88 & APMUKS(BJ) B221035.05-223249.1 & $78\arcsec$ from optical cluster \\
89 & ANTI-LEONID:[CME2001] 24 & X-ray source: $12\arcsec$ from optical cluster position \\
90 & APMUKS(BJ) B221513.16+000716.2 & galaxy: $24\arcsec$ from optical source \\
91 & CFRS 22.0770 & galaxy ($z=0.818800$): $12\arcsec$ from optical source \\
92 & SDSS J221833.05+001737.3 & galaxy ($z=0.332172$):  $0.0\arcsec$ from opt. src.\\
93 & NVSS J223948+033801 & radio source: $102\arcsec$ from optical centroid \\
94 & 2MASX J22395403+0332457 & galaxy: $78\arcsec$ from optical centroid \\
95 & 2MASX J22402529+0333046 & galaxy: $108\arcsec$ from optical centroid \\
96 & 2MASX J22402529+0333046 & galaxy: $54\arcsec$ from optical centroid \\
97 & [B2002a] 36 & galaxy cluster: $102\arcsec$ from optical cluster position \\
98 & NVSS J230312+085012 & radio source: $66\arcsec$ from optical position \\
99 & CXOSEXSI J234758.8+010344 & X-ray source: $6\arcsec$ from optical centroid \\
100 & CXOSEXSI J234815.8+005351 & X-ray source: $6\arcsec$ from optical centroid \\
101 & CXOSEXSI J234839.5+010828 & X-ray source: $12\arcsec$ from optical position \\
102 & APMUKS(BJ) B234658.77+004331.4 & galaxy: $54\arcsec$ from optical cluster \\
\enddata
\tablenotetext{a}{Published references acquired from the NASA/IPAC 
Extragalactic Database.}
\tablenotetext{b}{Blank field if published sources $>2\arcmin$ from object position.}
\end{deluxetable}

\clearpage

\clearpage

\clearpage

\clearpage

\clearpage

\clearpage

\clearpage
\clearpage

\end{document}